\documentclass[useAMS,usenatbib]{mnras}
\usepackage{graphicx}
\usepackage{amsmath}
\usepackage{amssymb}
\usepackage[TABTOPCAP]{subfigure}
\usepackage{url}
\usepackage{multirow}
\usepackage{grffile}

\newcommand{\m}{\rm\thinspace m}

\newcommand{\msq}{\hbox{$\m^2\,$}}

\newcommand{\s}{\rm\thinspace s}
\newcommand{\ks}{\rm\thinspace ks}

\newcommand{\Hz}{\rm\thinspace Hz}

\newcommand{\Msun}{\hbox{$\rm\thinspace M_{\odot}$}}

\newcommand{\keV}{\rm\thinspace keV}

\newcommand{\cts}{\rm\thinspace ct}
\newcommand{\ctsps}{\hbox{$\cts\s^{-1}\,$}}

\newcommand{\rg}{\rm\thinspace $r_\mathrm{g}$}

\title[X-ray timing with Gaussian processes]{Low frequency X-ray timing with Gaussian processes and reverberation in the radio-loud AGN 3C\,120}
\author[D. R. Wilkins]{D. R. Wilkins\thanks{E-mail: dan.wilkins@stanford.edu}\thanks{Einstein Fellow}\\
Kavli Institute for Particle Astrophysics and Cosmology, Stanford University, 452 Lomita Mall, Stanford, CA 94305, USA \\
}
\begin{document}

\date{Accepted 2019 August 12. Received 2019 August 09; in original form 2019 July 14}

\pagerange{\pageref{firstpage}--\pageref{lastpage}} \pubyear{2019}

\maketitle

\label{firstpage}

\begin{abstract}
A framework is developed to perform Fourier-domain timing analysis on X-ray light curves with gaps, employing Gaussian processes to model the probability distribution underlying the observed time series from which continuous samples can be drawn. A technique is developed to measure X-ray reverberation from the inner regions of accretion discs around black holes in the low frequency components of the variability, on timescales longer than can be probed employing standard Fourier techniques. This enables X-ray reverberation experiments to be performed using data from satellites in low-Earth orbit such as \textit{NICER}, \textit{NuSTAR} and the proposed X-ray timing mission \textit{STROBE-X}, and enables long timescale reverberation around higher mass AGN to be measured by combining observations. Gaussian processes are applied to observations of the broad line radio galaxy 3C\,120 spanning two orbits with \textit{XMM-Newton} to measure the relative time lags of successive X-ray energy bands. The lag-energy spectrum between $5\times 10^{-6}$ and $3 \times 10^{-5}$\Hz, estimated using Gaussian processes, reveals X-ray reverberation from the inner accretion disc for the first time in this radio-loud AGN. Time lags in the relativistically broadened iron K line are significantly detected. The core of the line lags behind the continuum by $(3800\pm1500)$\s, suggesting a scale height of the corona of $(13 \pm 8)$\rg\ above the disc. The ability to compare the structure of coronae in radio loud AGN to their radio quiet counterparts will yield important insight into the mechanisms by which black holes are able to launch jets.

\end{abstract}

\begin{keywords}
accretion, accretion discs -- black hole physics -- galaxies: active -- methods: statistical -- relativistic processes -- X-rays: galaxies.
\end{keywords}

\section{Introduction}
The discovery of X-ray reverberation off of the inner regions of the accretion discs around supermassive black holes in active galactic nuclei (AGN) has recently paved the way to great advances in understanding the extreme environments around black holes, the accretion process, and the mechanism by which supermassive black holes are able to power some of the most luminous objects in the Universe.

Continuum X-rays emitted from a putative corona of energetic particles in the immediate vicinity of the black hole irradiate the accretion disc. This produces a characteristic reflection spectrum through Compton scattering, thermal bremsstrahlung emission, photoelectric absorption and subsequent fluorescent line emission \citep{george_fabian, ross_fabian}. While the fluorescent lines are narrow in the rest frame of the emitting material, the combination of Doppler shifts, due to the orbital motion of the disc, and strong gravitational redshifts close to the black hole result in these lines appearing broadened, with a blueshifted wing and a redshifted tail extending to low energies \citep{fabian+89}. Most notably, a prominent, relativistically broadened, iron K$\alpha$ line is observed around its rest frame energy of 6.4\keV\ \citep{matt+97}, while the broadening of iron L, oxygen, nitrogen and other emission lines below 1\keV\ causes these features to become blended together to form at least part of a soft excess that is detected above the power law continuum spectrum.

X-ray reverberation is observed where variability in X-ray energy bands dominated by reprocessed emission from the accretion disc is seen to lag behind correlated variations in energy bands dominated by the primary continuum. Such time lags are seen in the soft X-ray excess below 1\keV\ \citep{fabian+09}, in the iron K line around 6.4\keV\ \citep{zoghbi+2012,zoghbi+2013,kara+13} and in the Compton hump at 20\keV\ \citep{zoghbi+2014,kara+2015} and are attributed to the additional light travel time between the primary X-ray source and the reprocessing disc. X-ray reverberation has been detected in more than 20 supermassive black holes. The lag is seen to increase as a function of black hole mass \citep{demarco+2012,kara_global}, demonstrating the relationship between the light travel time and the characteristic scale length in the gravitational field around the black hole. Lag times are short, corresponding to the light travel time across between 2 and 10\rg\ (where 1\rg$=GM/c^2$, the radius of the event horizon of a maximally spinning black hole), showing that X-ray reverberation is indeed probing structures in the immediate vicinity of the event horizon and that the corona producing the X-ray continuum is compact \citep{lag_spectra_paper,cackett_ngc4151}. It has also been argued that short timescale lags can be explained by reprocessing from more distant material ($\sim 100$\rg\ from the central engine) if the time lags are diluted by a strong, directly-observed continuum component \citep{miller+10,mizumoto+2018}. Detailed joint analysis of the X-ray spectrum, variability and the energy dependence of the time lags is requried to break this degeneracy.

Time lags between X-ray energy bands are measured from the Fourier transforms of the light curves \citep{reverb_review} and are measured as a function of Fourier frequency. In effect, this is dividing up the stochastic variability of the source into slowly (low frequency) and rapidly (high frequency) varying components. X-ray reverberation from the accretion disc is typically observed in the higher frequency variability components, where the soft excess and iron K line are seen to lag behind the continuum-dominated 1-4\keV\ band. At lower frequencies, the reverberation signal is not seen and the lag is seen to increase smoothly as a function of energy. This delay in response time of higher energy X-rays is attributed to variability within the continuum itself and is thought to originate from the propagation of fluctuations through the corona itself \citep{miyamoto+89,arevalo+2006}. Moreover, the highest frequency at which a reverberation signal can be detected depends upon the intrinsic lag time (and hence the scale-height of the corona above the accretion disc). For a lag time of $\tau$, the reverberation signal is not seen in frequency components above $f\sim 1/2\tau$, at which point the delay causes the phase difference between the two signals to wrap around from $\pi$ to $-\pi$ \citep{lag_spectra_paper}. The frequency at which the reverberation signal is detected is found to decrease with increasing black hole mass and magnitude of the lag \citep{kara_global}.

By observing the evolution with Fourier frequency of the lag structure as a function of X-ray energy, \citet{1zw1_corona_paper} were able to discover evidence of structure and time-evolution within the X-ray emitting corona. Combining the pattern of time delays as a function of energy over the redshifted wing of the iron K line (in which lower energy photons are emitted from regions of the disc closer to the black hole) with information about the extent of the corona gained from the time-averaged profile of the emission line in the X-ray spectrum \citep{understanding_emis_paper}, a picture is emerging of a bright, jet-like collimated core that drives the rapid X-ray variability embedded within a more slowly varying component of the corona extending over the surface of the innermost parts of the accretion disc \citep{propagating_lag_paper}.

X-ray reverberation is predominantly observed in radio-quiet AGN. The discovery of a collimated core within the corona raises the tantalising question of how the corona may be related to the large-scale jets that are seen in radio galaxies and radio-loud AGN; namely, is this core the base of a (failed) jet? Is the core of the corona produced if the energy in what would form the jet is dissipated in the magnetosphere close to the black hole? Or are the two structures unrelated? In order to address these questions, it is important to measure the structure of the corona in radio-loud AGN to understand the similarities and differences that could shed light on the mystery of why it is some black holes launch jets while others do not.

3C\,120 is a nearby ($z=0.033$), bright, broad line radio galaxy (BLRG) that exhibits a one-sided super-luminal jet. Unlike many radio galaxies, however, 3C\,120 shows the reflection of an X-ray continuum from the inner regions of the accretion disc, with the detection of a relativistically broadened iron K$\alpha$  line in spectra observed by \textit{Suzaku} \citep{kataoka+2007,cowperthwaite+2012}. 3C\,120 exhibits a prominent jet cycle \citep{marscher+2002} and conducting detailed multi-epoch X-ray studies with \textit{Suzaku} and \textit{XMM-Newton}, \citet{lohfink_3c120} were able to observe evidence of the interplay between the accretion disc and the jet. They find that the X-ray spectrum is well described by a composite model consisting of jet emission, the X-ray continuum and reflection from the accretion disc, and that the ejection of knots within the jet coincides with the inner disc becoming truncated (and potentially being ejected) before refilling.

The detection of relativistically broadened reflecion components within the X-ray spectrum of 3C\,120 makes it the ideal target for X-ray reverberation studies among radio-loud AGN. Radio-loud AGN are typically supposed to harbour more massive black holes than their radio quiet counterparts. Indeed, optical reverberation mapping constrains the black hole mass in 3C\,120 to be $(5.7 \pm 2.7)\times 10^7$\Msun\ \citep{pozonunez+2012}, approximately an order of magnitude larger than the radio quiet AGN in which reverberation is measured. Reverberation signatures are therefore expected to be found on much longer timescales and at approximately an order of magnitude lower in Fourier frequency than in the radio quiet sample.

The lowest frequencies at which X-ray timing analysis can be conducted using the standard Fourier transform techniques is limited by the longest continuous light curve segments that are available. Where X-ray observations are conducted by satellites, this is limited by the orbital period of the satellite when the target becomes occulted by the Earth during each revolution. To conduct timing analyses and measure X-ray reverberation at lower frequencies, it is therefore necessary to combine segments of the time series from successive orbits, with gaps. This problem is more pronounced when observations are made with X-ray observatories in low-Earth orbit (LEO) such as \textit{Suzaku}, \textit{NuSTAR} or \textit{NICER}, where orbital gaps every 90 minutes place the lower frequnecy limit at $\sim 10^{-4}$\Hz.

To overcome this limitation, \citet{zoghbi_gap} develop a technique to calculate time lags from unevenly sampled light curves (\textit{i.e.} light curves with gaps), based upon the method of  \citet{bond_powerspec} to estimate the power spectrum of the cosmic microwave background or CMB \citep[see also][]{miller_gaps}. A model is fit to the pair of light curves that simultaneously describes the autocorrelation of each light curve and the cross-correlation between the two. This model encodes the time lags that are to be measured and can be quite general, parameterised in terms of the cross power and phase lag at a set of Fourier frequencies. This reproduces what would be measured by the Fourier transform, although data are fit in the time domain. It is through this technique that it was possible to measure reverberation time lags in the Compton hump using \textit{NuSTAR}.

The method of \citet{zoghbi_gap} employs a form of Gaussian process to compute the likelihood function that is maximised to fit the power spectrum and lags to the data. Gaussian processes offer a flexible framework to define sampling distributions of continuous functions \citep{gpbook}, such as the stochastic X-ray light curves of accreting black holes, in a purely data-driven manner. Rather than applying a model of the power spectrum and lags to the observed light curves, a Gaussian process model can be constructed of the probability distribution underlying the observed light curves. Continuous light curves can then be drawn from this distribution to circumvent gaps in the light curves and analyse uncertainties in a probabilistic manner. Indeed, \citet{reynolds_mcg6_reverb} conducted a simplified Gaussian process analysis on light curves of MCG--6-30-15 recorded by \textit{RXTE}, building an \textit{optimal reconstruction} to estimate the light curves in the gaps using a model for the covariance. We here seek to build on that technique to construct a flexible X-ray timing framework.

There have recently been rapid developments in computational routines for Gaussian process analysis that have, in many ways, come with recent advances in machine learning (Gaussian processes are often described as a step towards machine learning models). \citet{czekala_gp} successfully apply Gaussian processes to the variability of complex stellar spectra for the precision measurement of radial velocity variations in close stellar binaries and for exoplanet detection.

In this article, a framework is developed for conducting X-ray timing analyses in the Fourier domain using Gaussian processes. The applicability of this framework to the measurement of X-ray reverberation at low frequencies across gaps in light curves is tested and the framework is applied to measure X-ray reverberation for the first time in the BLRG 3C\,120.

\section{Gaussian processes}
In order to perform timing analyses, we wish to obtain the underlying continuous light curves without gaps between spacecraft orbits or observation segments. In order to do this, the underlying time series is modelled as a \textit{Gaussian process} which is fit to the observed data points and then used to generate realisations of the underlying time series at times no data are available \citep[see also][]{gpbook}. 

Composing a data vector, $\mathbf{d}$, from the observed count rate or flux in each time bin, with a corresponding input vector, $\mathbf{t}$, composed from the time of each bin.
\begin{align}
\mathbf{t} &= 
\begin{pmatrix}
t_1 \\ t_2 \\ \vdots \\ t_N
\end{pmatrix}
&
\mathbf{d} &= 
\begin{pmatrix}
d_1 \\ d_2 \\ \vdots \\d_N
\end{pmatrix}
\end{align}
For the time series to be a Gaussian process, the data vector, $\boldsymbol{d}$, is drawn from a multivariate Gaussian distribution and the observed light curve is one realisation of the Gaussian process. Let $\boldsymbol{x}$ be a Gaussian random vector with mean vector $\boldsymbol{\mu}_x$ and covariance matrix $\boldsymbol{\Sigma}_{xx}$ between pairs of elements in $\boldsymbol{d}$.
\begin{align}
\mathbf{x} \sim \mathcal{N}(\boldsymbol{\mu}_x, \boldsymbol{\Sigma}_{xx})
\end{align}
The likelihood function of $\boldsymbol{x}$ is
\begin{align}
\label{likelihood.equ}
\begin{split}
p(\boldsymbol{x} | &\boldsymbol{\mu}_x, \boldsymbol{\Sigma}_{xx}) \\ &= \frac{1}{(2\pi)^N |\boldsymbol{\Sigma}_{xx}|} \exp\left(-\frac{1}{2}(\boldsymbol{x} - \boldsymbol{\mu}_x)^T \boldsymbol{\Sigma}_{xx}^{-1} (\boldsymbol{x} - \boldsymbol{\mu}_x) \right)
\end{split}
\end{align}
The time series is assumed to be stationary with a constant mean value over the course of the observations, hence $\boldsymbol{\mu}_x$ is replaced by the mean count rate or flux and variability in the time series is generated by the realisation of each element within the Gaussian process data vector. The elements of the covariance matrix are generated by a \textit{kernel funtion}, $k(t_1, t_2)$, between pairs of input values, which will have one or more variable parameters, so-called \textit{hyperparameters}, that describe the variability in the time series.

The values of the hyperparameters are fit to the observed data points. The posterior probability distribution of these parameters is specified by the marginal likelihood, Equation~\ref{likelihood.equ}. The hyperparameters are optimised by maximising the likelihood.

Once the hyperparameters have been optimised using the observed data $\boldsymbol{d}$ at times $\boldsymbol{t}$, realisations of the Gaussian process can be generated over a different set of time bins, $\boldsymbol{t}_*$ (\textit{i.e.} including the gaps in the observation).

If $\boldsymbol{x}$ and $\boldsymbol{y}$ are jointly Gaussian random vectors drawn from the multivariate Gaussian distribution
\begin{align}
\begin{pmatrix}
\boldsymbol{x} \\ \boldsymbol{y}
\end{pmatrix}
= \mathcal{N}\left(
\begin{pmatrix}
\boldsymbol{\mu}_x \\ \boldsymbol{\mu}_y
\end{pmatrix},
\begin{pmatrix}
\boldsymbol{\Sigma}_{xx} & \boldsymbol{\Sigma}_{xy} \\ \boldsymbol{\Sigma}_{yx} & \boldsymbol{\Sigma}_{yy}
\end{pmatrix}
\right)
\end{align}
the conditional distribution of $\boldsymbol{x}$ given $\boldsymbol{y}$ is the normal distribution
\begin{align}
\label{conditional.equ}
(\boldsymbol{x} | \boldsymbol{y}) \sim \mathcal{N}(\boldsymbol{\mu}_x + \boldsymbol{\Sigma}_{xy}\boldsymbol{\Sigma}_{yy}^{-1}(\boldsymbol{y} - \boldsymbol{\mu}_y), \boldsymbol{\Sigma}_{xx} - \boldsymbol{\Sigma}_{xy}\boldsymbol{\Sigma}_{yy}^{-1}\boldsymbol{\Sigma}_{yx})
\end{align}
The covariance matrix is composed of the submatrices $\boldsymbol{\Sigma}_{xx}$ and $\boldsymbol{\Sigma}_{yy}$, the covariances between elements of $\boldsymbol{x}$ and $\boldsymbol{y}$ as well as $\boldsymbol{\Sigma}_{xy}$ and $\boldsymbol{\Sigma}_{yx} = \boldsymbol{\Sigma}_{xy}^T$ corresponding to the covariances between elements of $\boldsymbol{x}$ and $\boldsymbol{y}$ (which are zero if the two data series are independent)

The Gaussian process, with optimised hyperparameters, describing the variability between data points within one realisation, $\boldsymbol{d}_*$, is anchored to the observed data points, $\boldsymbol{d}$ by setting $\boldsymbol{x} = \boldsymbol{d}_*$ and $\boldsymbol{y} = \boldsymbol{d}$. The cross-covariance submatrix, $\boldsymbol{\Sigma}_{xy}$ is constructed by evaluating the kernel function between the appropriate pairs of time bins between the two series $\boldsymbol{t}_*$ and $\boldsymbol{t}$ since we wish to draw $\boldsymbol{d}_*$ from the same distribution as $\boldsymbol{d}$ with the same kernel and hyperparameters.

The Gaussian process realisation of the underlying time series is generated by taking random draws from the condition distribution $(\boldsymbol{d}_* | \boldsymbol{d})$ in Equation~\ref{conditional.equ}. At times in $\boldsymbol{t}$ at which real data are observed, random draws of $\boldsymbol{d}_*$ will automatically follow the corresponding points in $\boldsymbol{d}$.

We use the Gaussian process framework in the Python package \textsc{scikit-learn}\footnote{\url{http://scikit-learn.org}} implemented for application to light curves and X-ray timing analysis in the package \textsc{pylag}\footnote{\url{http://github.com/wilkinsdr/pylag}}.

\section{Building the kernel function}
It is necessary to select a form for the kernel function that accurately describes the variability of the time series such that the count rate or flux can be predicted in the gaps between observations.

X-ray light curves of AGN are described by broken power law power spectral densities (PSDs). The PSD follows approximately $f^{-1}$ up to a break frequency $f_\mathrm{br}$ that scales with the mass of the black hole, then falls off as $f^{-2}$ at high frequencies \citep{mchardy+2004,uttley_mchardy-2005,mchardy+2006}. The break frequency is typically $\sim 10^{-5}$\Hz\ and time delays corresponding to X-ray reverberation from the inner regions of the accretion disc are seen in the high frequency portion of the power spectrum around $10^{-4} \sim 10^{-3}$\Hz\ where the PSD follows $f^{-2}$.

An observation was simulated by generating a random light curve with the specified PSD using the algorithm of \citet{timmer_konig}. Once the full light curve was generated, data points were removed from periodic segments to simulate orbital gaps. For the initial assessment of kernel functions, we consider the case of a satellite in low Earth orbit such as \textit{NuSTAR} with gaps 2400\s\ long beginning every 5700\s.

The autocovariance of the light curve is entirely specified by the PSD, hence it is useful to compare the analytic kernel function to the expected autocovariance. From Parseval's theorem, the autocovariance and the PSD are Fourier transform pairs, thus a simple Fourier transform converts between the time domain covariance, as modelled by the kernel function, and the frequency domain PSD.

\subsection{Describing the variability}
We seek simple analytic kernel functions that can be computed efficiently each time the hyperparameters are optimised and draws are taken from the Gaussian process. We shall consider analytic kernel functions that are commonly employed in Gaussian process analyses and are implemented within \textsc{scikit-learn} to assess their applicability to X-ray light curves of accreting black holes.

\subsubsection{Squared exponential}
The most common starting point in Gaussian process analysis is the \textit{squared exponential} kernel function (also known as the \textit{radial-basis function, RBF} kernel). In this case
\begin{align}
k(t_i, t_j | \sigma, l) = \sigma^2 \exp\left[-\frac{1}{2}\left(\frac{t_j - t_i}{l} \right)^2 \right]
\end{align}
This kernel has just two free parameters, a normalisation, $\sigma^2$, to fit the variance of the time series and a characteristic length scale of the variations, $l$.

\begin{figure*}
\centering
\subfigure[Light curve] {
\includegraphics[height=49mm]{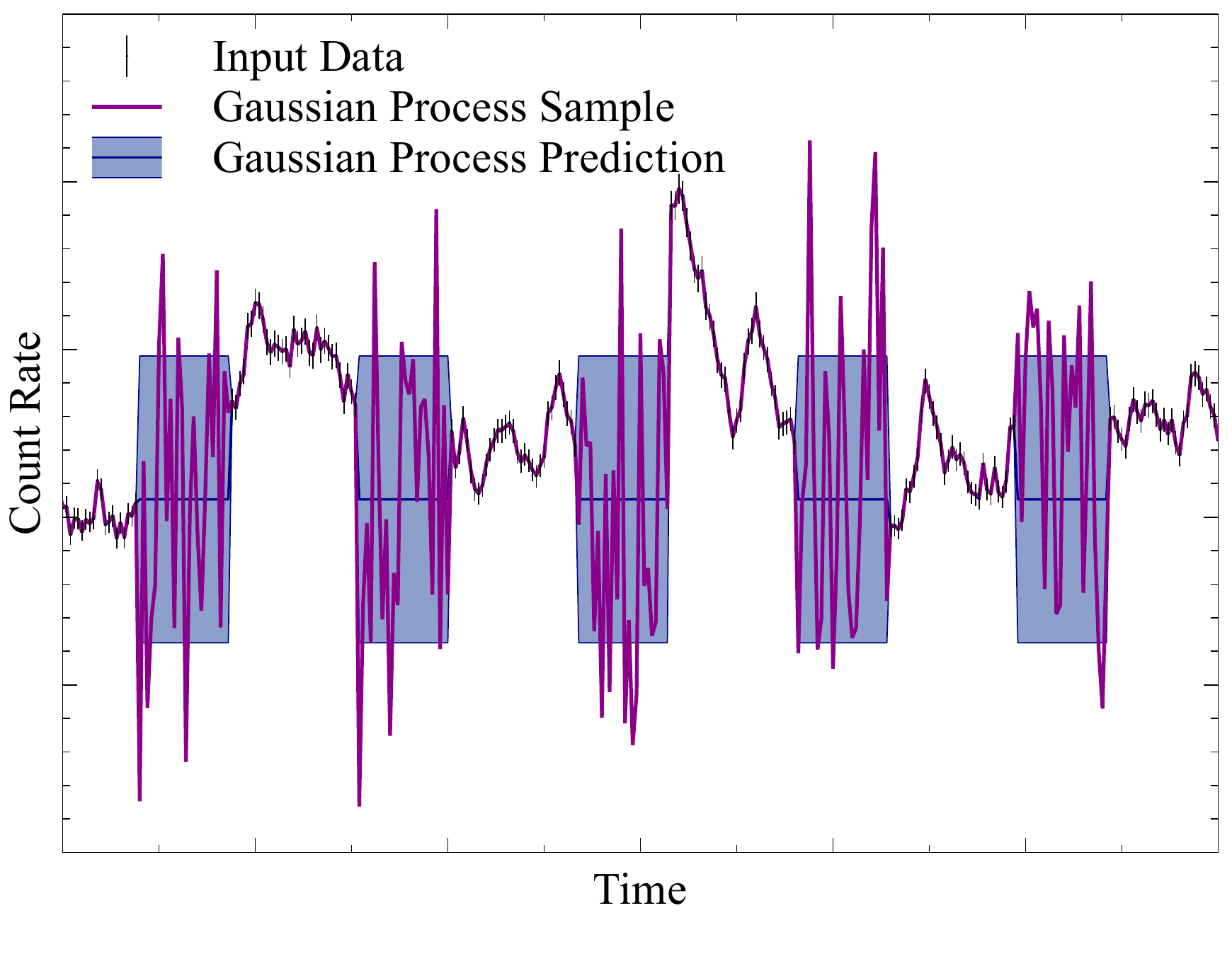}
\label{se_kernel.fig:lc}
}
\subfigure[Autocorrelation, power law PSD] {
\includegraphics[height=49mm]{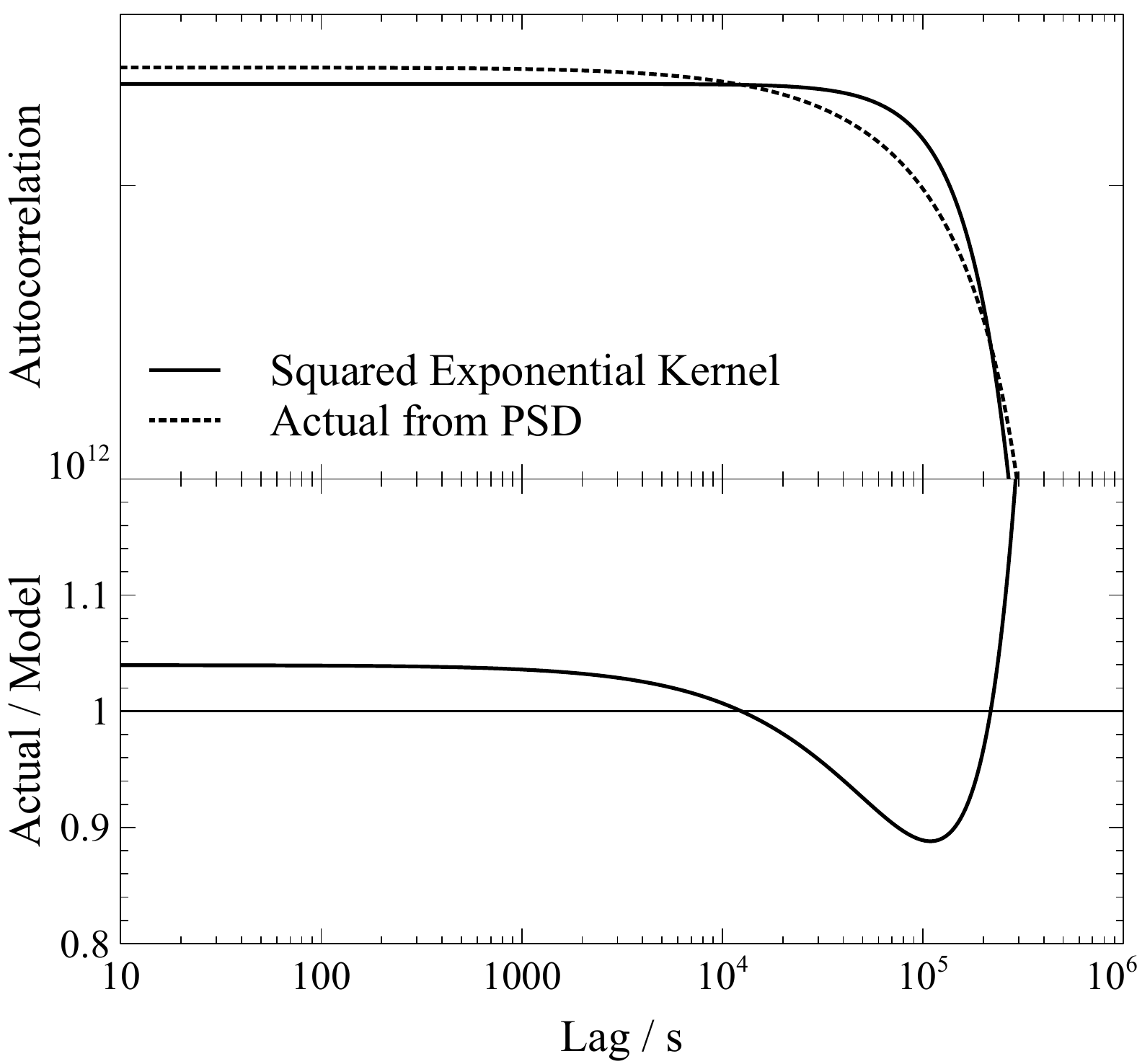}
\label{se_kernel.fig:psd}
}
\subfigure[Autocorrelation, broken power law] {
\includegraphics[height=49mm]{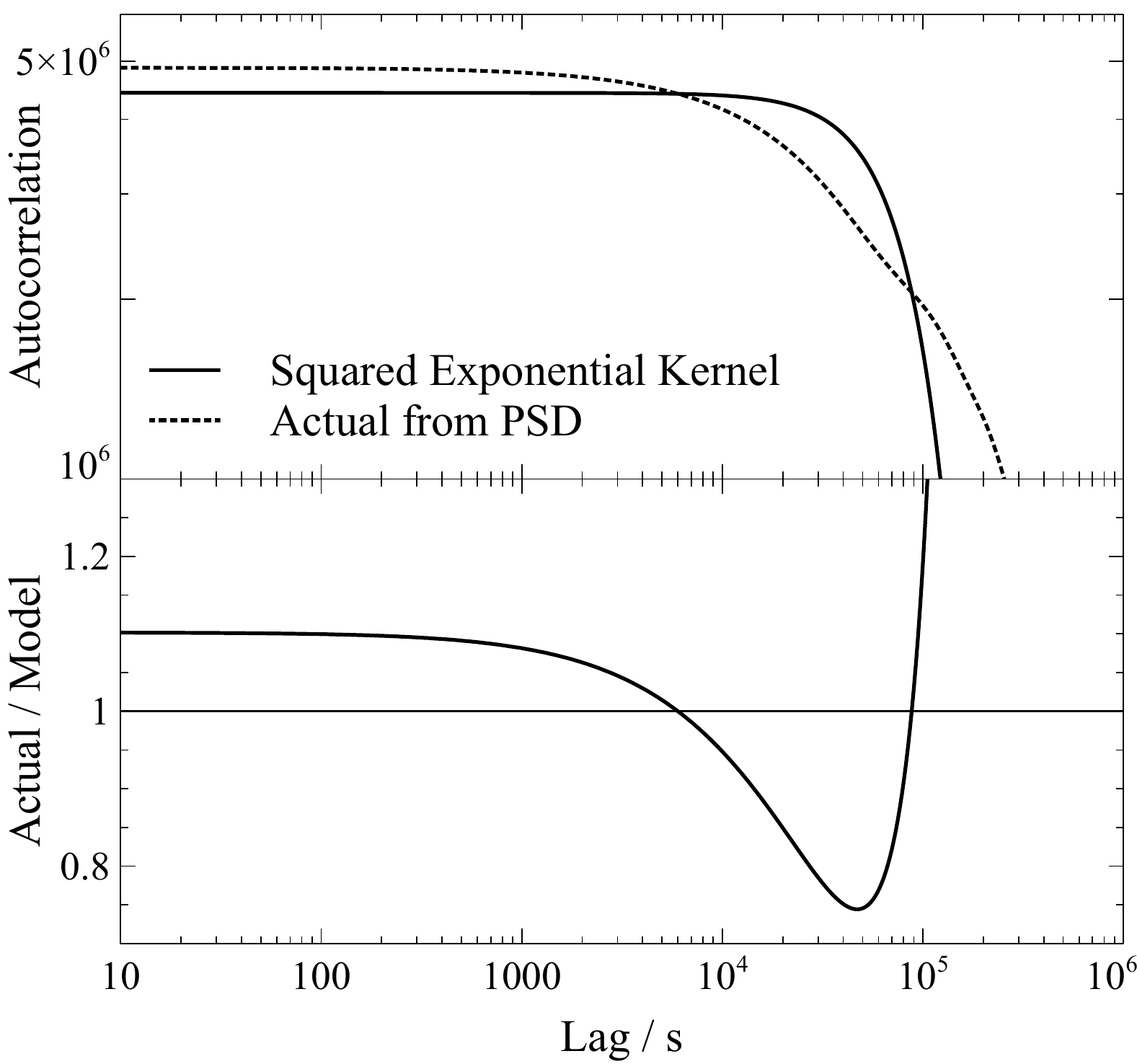}
\label{se_kernel.fig:brokenpl_psd}
}
\caption[]{Assessment of the squared exponential kernel in describing the variability in an X-ray light curve of an accreting black hole. \subref{se_kernel.fig:lc} The simulated light curve with a $f^{-2}$ power spectrum and gaps corresponding to observation by a satellite in low-Earth orbit, with a single sample drawn from the Gaussian process shown by the purple line and the prediction from averaging 1000 samples shown by the blue band. \subref{se_kernel.fig:psd} Comparison of the best-fitting kernel function to the true autocorrelation function for a single power law $f^{-2}$ PSD. \subref{se_kernel.fig:brokenpl_psd} Comparison of the kernel function to the true autocorrelation for a broken power law PSD breaking from $f^{-1}$ to $f^{-2}$ at $10^{-5}$\Hz.}
\label{se_kernel.fig}
\end{figure*}

Fig.~\ref{se_kernel.fig:lc} shows an example random time series with a power spectrum following $f^{-2}$ and gaps equivalent to those present in astronomical light curves observed from low Earth orbit. Samples are drawn from a Gaussian process with a squared exponential kernel function. A single sample is shown by the purple line while the blue shaded region shows the predicted value (and associated standard deviation) of the time series in the gap, obtained by averaging over 1000 samples.

While the squared exponential kernel reproduces the variability on the shortest timescales, it is clear that the longer-term variations are not reproduced. The predicted time series rapidly returns to the mean value of the entire measured series and remains at this same constant value over the course of each gap, rather than following the long-term trend from one side of the gap to the other.

The fidelity of the Gaussian process representation of the light curve can be further investigated by comparing the best-fitting kernel (\textit{i.e.} autocorrelation) function to the autocorrelation function of the original time series. Fig.~\ref{se_kernel.fig:psd} compares the best-fitting squared exponential kernel function to the true autocorrelation calculated from the Fourier transform of the PSD. The discrepancy between the shape of these functions can clearly be seen with the best-fitting function diverging rapidly from the true function at long timescales. In particular, the original time series has much more variability on long timescales than is predicted by the kernel. Lastly, Fig.~\ref{se_kernel.fig:brokenpl_psd} compares the kernel function to the autocorrelation function in the case of a random time series generated with a broken power law power spectrum, breaking from $f^{-1}$ to $f^{-2}$ at $10^{-5}$\Hz. Similar behaviour is seen to the case of a single power law power spectrum.

\subsubsection{Rational quadratic}
It is clear that the single timescale parameter in the squared exponential kernel function is insufficient to describe the timescales of variability in a time series with a typical power spectrum. One might therefore choose to sum squared exponential functions with different timescales, $l$. Summing an infinite series of squared exponential functions with scale lengths drawn from a gamma distribution yields the \textit{rational quadratic (RQ)} kernel:

\begin{align}
k(t_i, t_j | \sigma, l, \alpha) = \sigma^2 \left(1 + \frac{(t_j - t_i)^2}{2\alpha l^2} \right)^{-\alpha}
\end{align}

In this case, $\alpha$ is the scale mixture parameter and alters the relative contribution of different length scales, while $\sigma^2$, once again, is the normalisation.

\begin{figure*}
\centering
\subfigure[Light curve] {
\includegraphics[height=49mm]{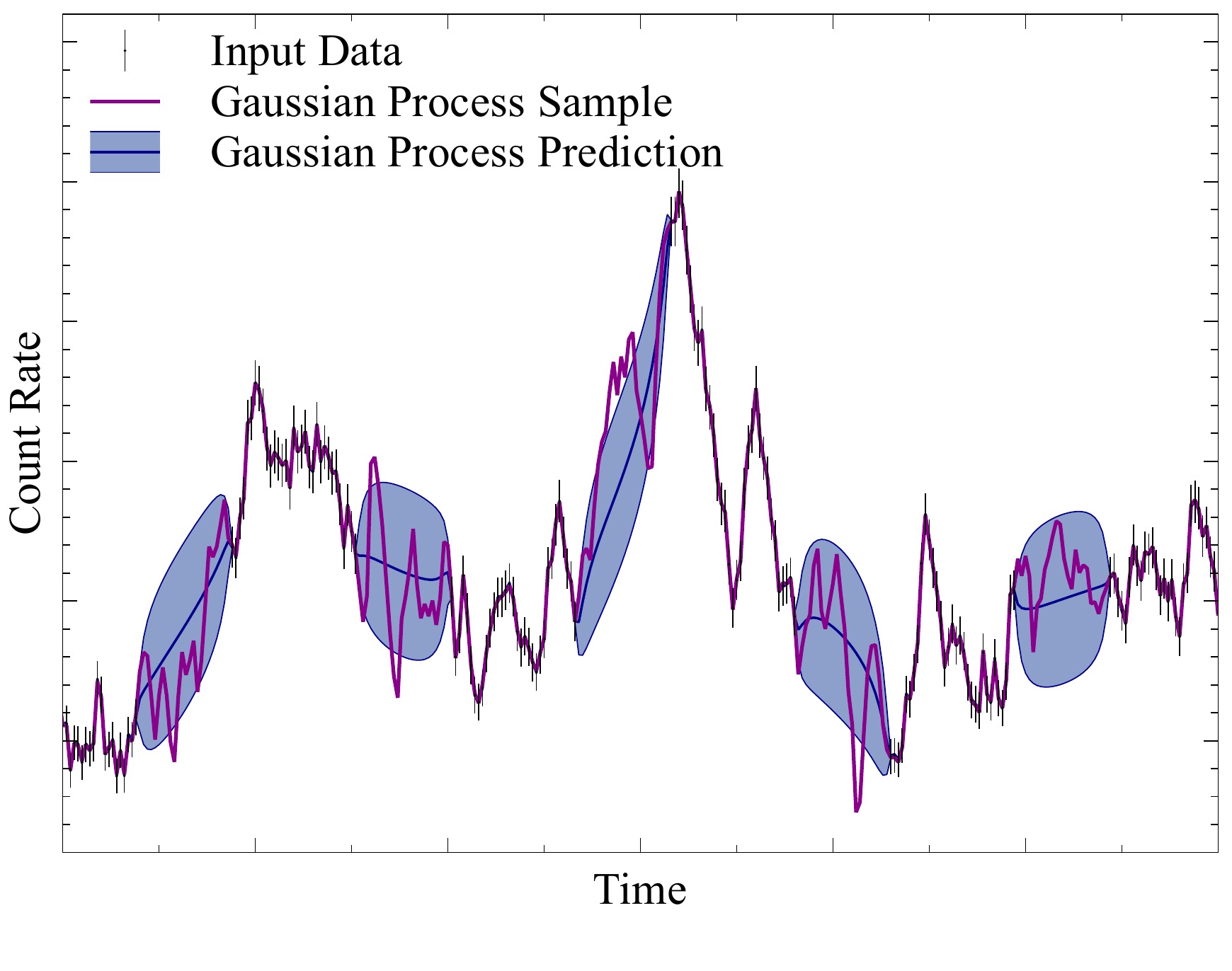}
\label{rq_kernel.fig:lc}
}
\subfigure[Autocorrelation, power law PSD] {
\includegraphics[height=49mm]{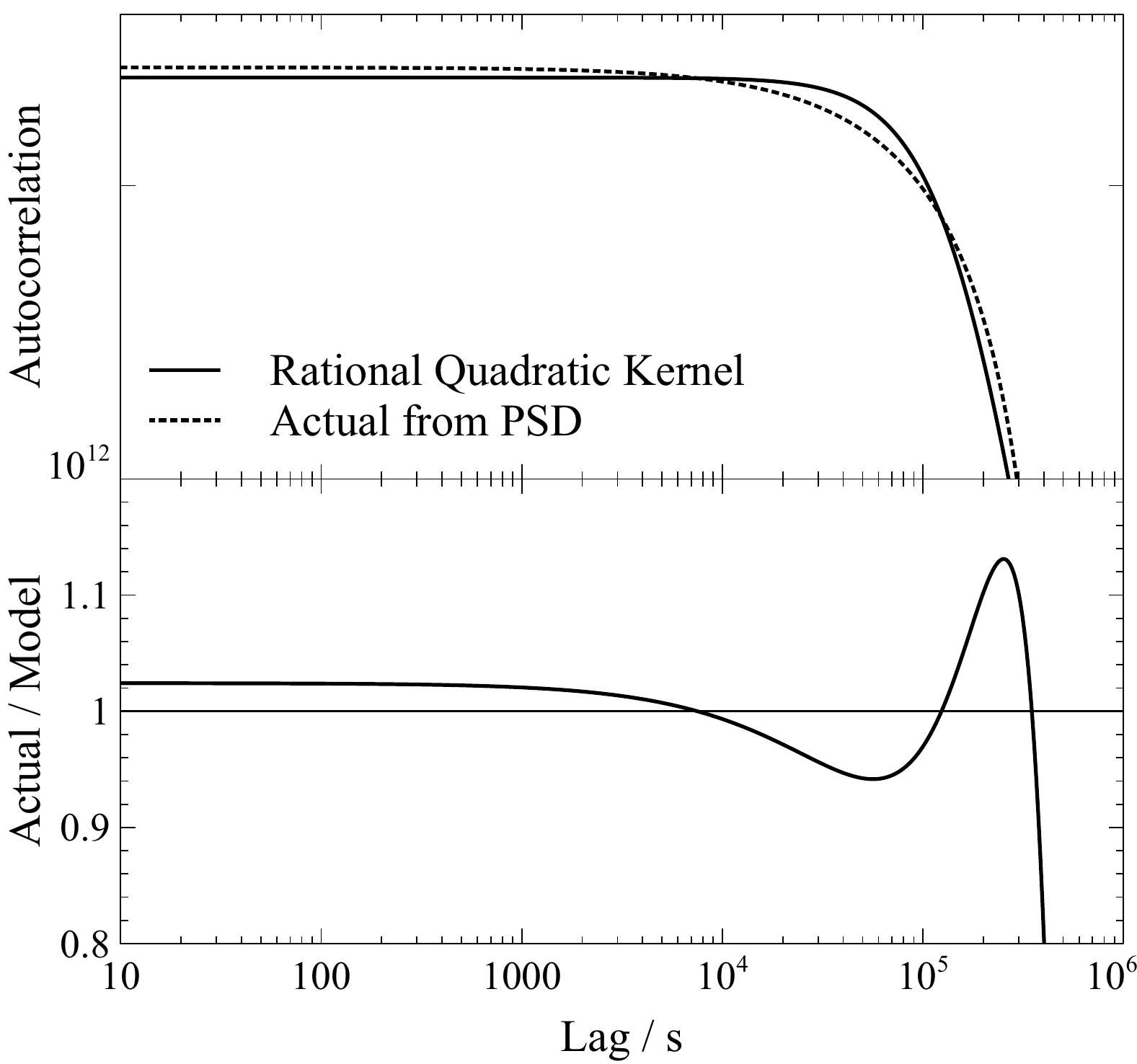}
\label{rq_kernel.fig:psd}
}
\subfigure[Autocorrelation, broken power law] {
\includegraphics[height=49mm]{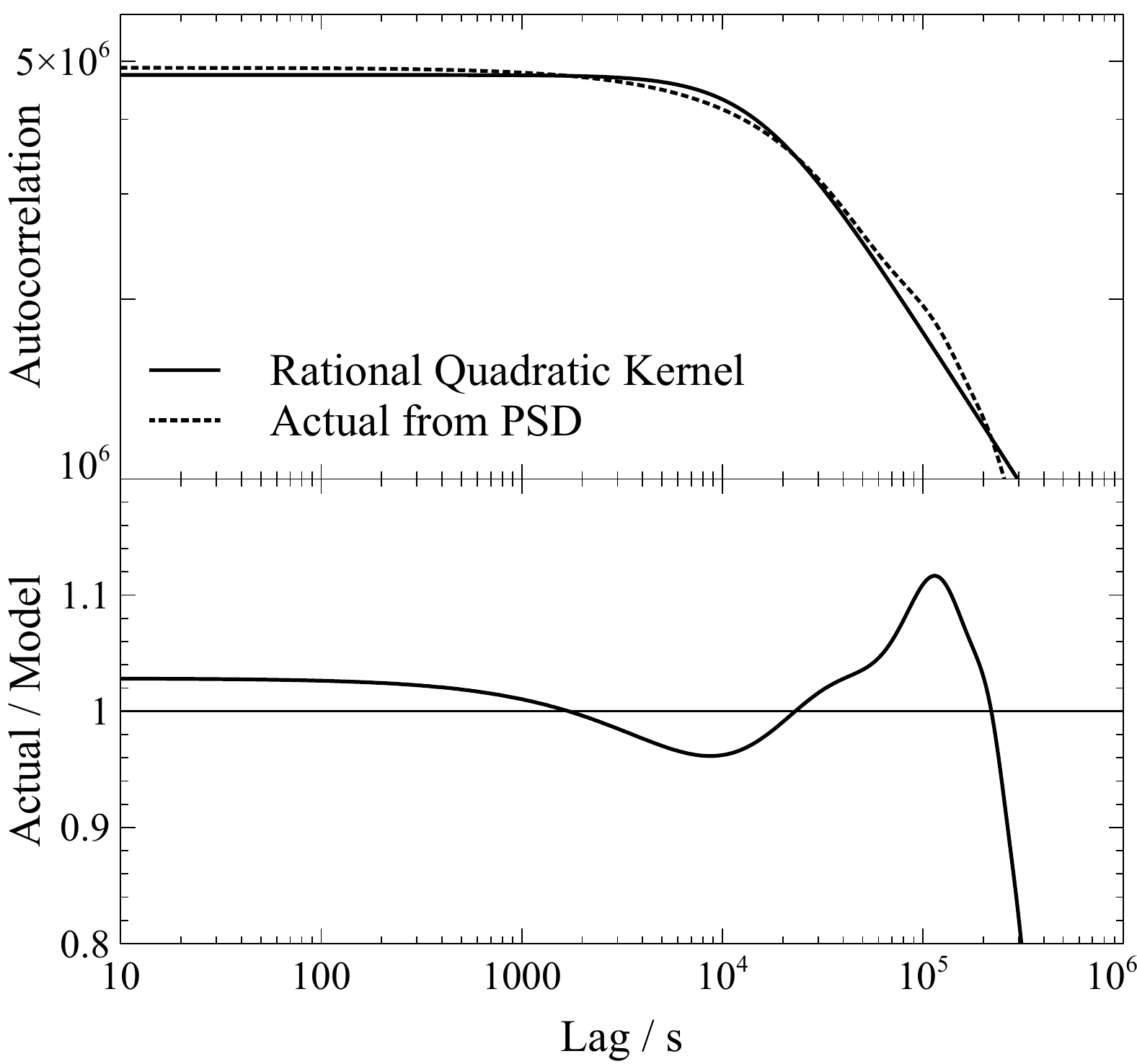}
\label{rq_kernel.fig:brokenpl_psd}
}
\caption[]{As Fig.~\ref{se_kernel.fig}, assessing the representation of a typical black hole X-ray light curve by the rational quadratic kernel.}
\label{rq_kernel.fig}
\end{figure*}

Fig.~\ref{rq_kernel.fig} assesses the performance of the rational quadratic kernel in representing time series described by a single power law, $f^{-2}$, or broken power law PSD. Providing a mixture of characteristic length scales enables the Gaussian process to reproduce both the long and short timescale variability with the samples displaying high frequency variability while the prediction over the course of a gap represents the variation in observed values across the gap, rather than reverting to the long-term mean.

Comparing the best-fitting kernel function to the autocorrelation function shows that the rational quadratic kernel produces significantly smaller residuals to the actual autocorrelation function. The autocorrelation function is faithfully reproduced up to timescales (lags between data points) of around 5000\s, as is evident from the constant ratio between the actual and model functions. If gaps up to 5000\s\ are present in the time series, the normalisation would be fixed by the observed data points and the predicted values over the gap would follow the correct autocorrelation.

It is only necessary to use the kernel function to predict data points up to the end of the gaps, thus errors on timescales beyond the length of gaps do not impact the predictions. From $10^{-4}$ to $10^5$\s, the rational quadratic kernel leads to an error of around 5 per cent in the autocorrelation, then around 10 per cent between $10^5$ and $4\times 10^5$\s\ (but switching from an over-prediction to an under-prediction of the variability power). At the longest timescales, the error diverges. Beyond the length of the light curve, the autocorrelation cannot be well modelled, emphasising the difficulty of extrapolating as opposed to interpolating the time series.

Similar behaviour is observed with the broken power law power spectral density. In this case, the autocorrelation is well described, with a flat ratio between the actual and model functions up to 2000\s, approximately 3 per cent error up to $2\times 10^4$\s\ and 10 percent error up to $2\times 10^5$\s.

\subsubsection{Matern kernel}
The Matern kernel is based upon the squared exponential kernel with length scale $l$, but in this case an additional parameter, $\nu$, is introduced to adjust the `smoothness' of the function.

\begin{align}
\begin{split}
k(t_i, t_j | \nu, \sigma) = \sigma^2 &\frac{1}{\Gamma(\nu) 2^{\nu-1}} \left(\sqrt{2\nu}\frac{t_j - t_i}{l}\right)^\nu \\ &K_\nu\left(\sqrt{2\nu}\frac{t_j - t_i}{l}\right)
\end{split}
\end{align}

$K_\nu$ is the modified Bessel function of the second kind. For general values of $\nu$, the Matern function is computationally intensive due to the requirement to compute this Bessel function. For $\nu = \frac{1}{2}, \frac{3}{2}, \frac{5}{2}$, however, the function simplifies to analytic forms. In particular, for $\nu = \frac{1}{2}$, the Matern kernel simplifies to the absolute exponential (as opposed to the squared exponential) kernel:
\begin{align}
k(t_i, t_j | \sigma, l) = \sigma^2 \exp\left[-\frac{1}{2}\left(\frac{t_j - t_i}{l} \right) \right]
\end{align}

\begin{figure*}
\centering
\subfigure[Light curve] {
\includegraphics[height=49mm]{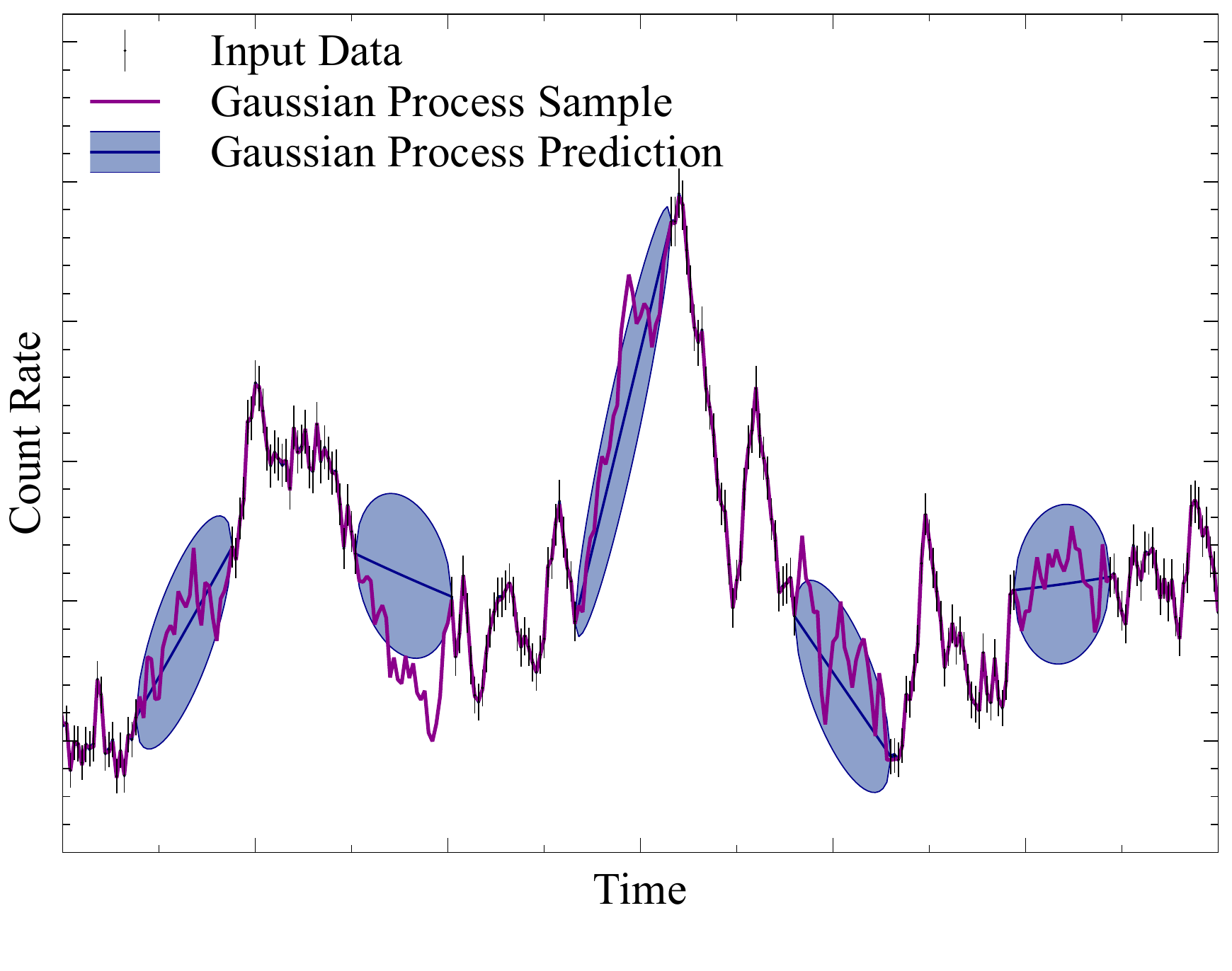}
\label{m1_kernel.fig:lc}
}
\subfigure[Autocorrelation, power law PSD] {
\includegraphics[height=49mm]{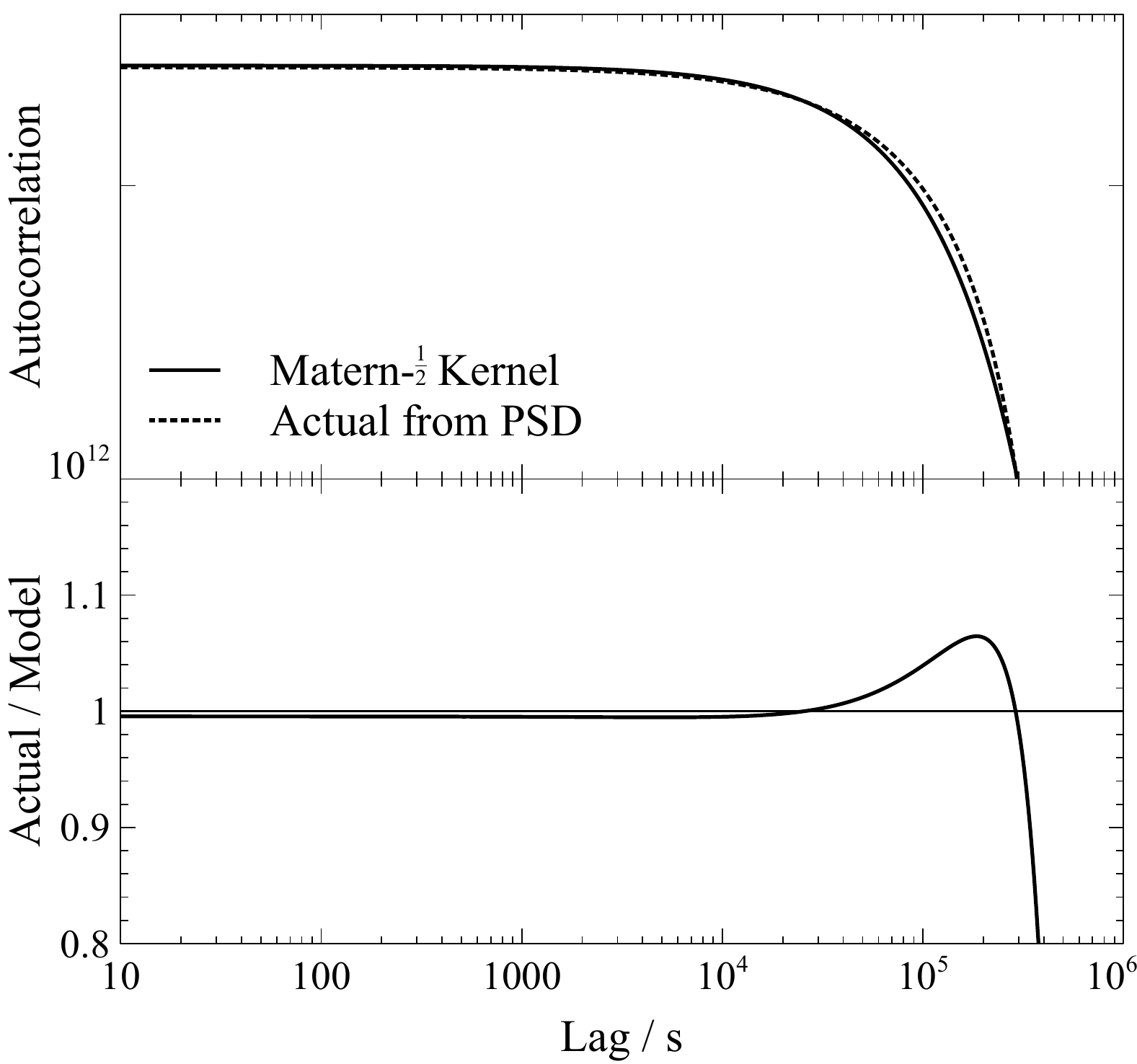}
\label{m1_kernel.fig:psd}
}
\subfigure[Autocorrelation, broken power law] {
\includegraphics[height=49mm]{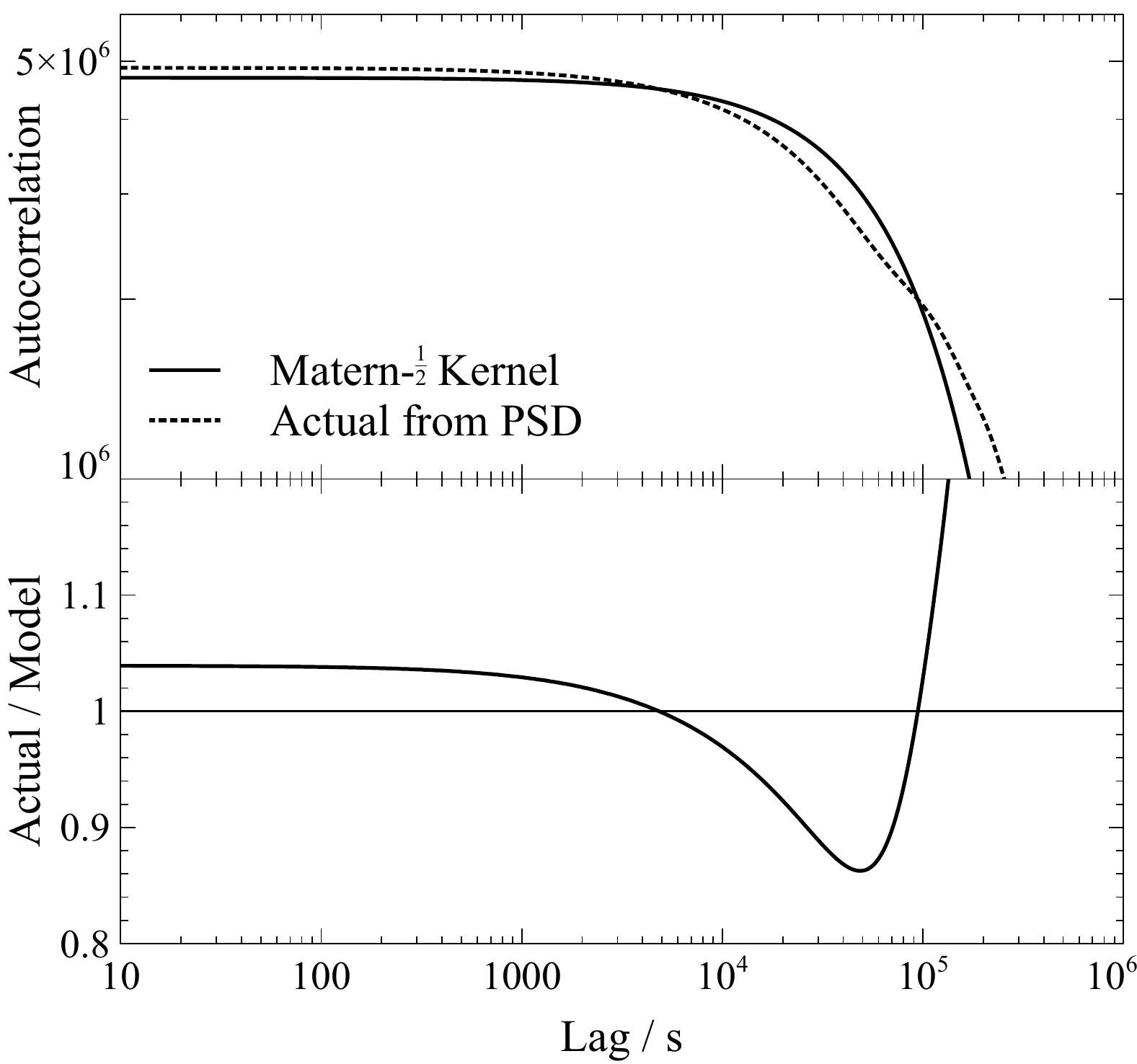}
\label{m1_kernel.fig:brokenpl_psd}
}
\caption[]{As Fig.~\ref{se_kernel.fig}, assessing the representation of a typical black hole X-ray light curve by the Matern-$\frac{1}{2}$ kernel.}
\label{m1_kernel.fig}
\end{figure*}

Comparing the time series predictions and best-fitting kernel funtions to the original time series and autocorrelation in Fig.~\ref{m1_kernel.fig} reveals that the Matern-$\frac{1}{2}$ kernel provides the most accurate description of the time series with the single power law, $f^{-2}$ PSD, maintaining an accurate representation of the autocorrelation up to timescales of $4 \times 10^{-4}$\s, then giving around 8 per cent error up to $4 \times 10^{5}$\s.

In the case of the broken power law PSD, the short timescale part of the autocorrelation function is reproduced accurately up to $5000$\s, reaching longer timescales than the rational quadratic kernel (where the constant ratio between the actual and model autocorrelation extended only up to 2000\s), but this is at the expense of the larger error of around 14 per cent at long timescales, up to $10^5$\s, at which point the error diverges.

\subsubsection{Selecting the kernel function}

From the selection of kernel functions that possess simple analytic forms that can be computed inexpensively, we find that the rational quadratic or Matern-$\frac{1}{2}$ kernel functions can be used to provide accurate Gaussian process models of time series possessing either a constant $f^{-2}$ power law or a broken power law PSD, depending upon the lengths of the gaps over which predictions are to be made.

In the case of a single power law PSD, the Matern-$\frac{1}{2}$ provides the most accurate description over the broadest range of timescales, with the representation of the autocorrelation function starting to become inaccurate over timescales longer than 5000\s. Errors as low as 8 per cent are maintained up to $4 \times 10^{5}$\s. This kernel is applicable to the light curves of AGN that are observed at timescales corresponding to frequencies well above the break frequency, allowing the PSD to be approximated as a single power law. 

Where the PSD is described by a broken power law, the choice of kernel function is less clear-cut. If predictions are required only over shorter timescales, a Matern-$\frac{1}{2}$ kernel function will provide accuracy over a greater range. The model autocorrelation function will enable predictions to be made over gaps of up to 5000\s, compared to only approximately 1000\s\ using the rational quadratic kernel. However, at longer timescales and for making predictions over longer gaps, the rational quadratic kernel provides significantly smaller error in the autocorrelation, so for gaps over $\sim 5000$\s\ would be more suitable.

So far, reproduction of the autocorrelation function, and hence the power spectrum, has been considered. In the next section, we shall explore the accuracy with which Gaussian processes can reproduce the phase information in time series and the impact of the errors manifested in the autocorrelation function.

\subsection{Noise}
While the intrinsic variability within a light curve, from the underlying physical processes driving the variation in luminosity, can be described in terms of analytic approximations to the autocorrelation function (the kernel function), the luminosity in each time bin will also vary due to any sources of noise in the observation.

The dominant source of noise is often Poisson noise associated with the arrival rate of photons. Poisson noise is uncorrelated between time bins. It can therefore be incorporated into the Gaussian process model by adding a diagonal term to the covariance matrix (a so-called \textit{white noise kernel}) equal to the contribution to the variance of each data point that is expected due to noise. The variance due to Poisson noise is equal to $N$, the photon count recorded in the time bin. Adding this diagonal term to the covariance matrix of the multivariate Gaussian essentially draws each data point from a Gaussian distribution corresponding to the noise around the distribution of the underlying variability. Of course, incorporating this into the Gaussian process assumes the noise follows a Gaussian distribution, \textit{i.e.} the high-count limit of the Poisson distribution applicable for greater than $\sim 20$ counts per bin.

Adding a white noise, diagonal kernel to model Poisson noise in the observation can fit the noise level to the observed data points, rather than needing to specify the level. This means that the best-fitting hyperparameter values can be marginalised over any uncertainty in the precise noise contibution. If the mean count rate remains approximately constant and the intrinsic variability dominates over the noise, a constant noise variance can be applied across all time bins.

If the observed light curve is sufficiently variable that the noise contribution varies from one time to another, independent noise variance values can be applied to each bin. In the latter case, however, care must be taken if fitting the noise as free hyperparameters if the noise values are unconstrained. Any variability not accounted for by the kernel function could be fit using a noise model that is too general, which would limit the ability of the Gaussian process to propagate the variability between the observed data points into the gaps to make predicitons.

\section{X-ray timing with Gaussian processes}
Once the Gaussian process has been optimised to the observed data points, continuous light curve samples can be drawn and each can be analysed employing any of the standard Fourier-domain spectral timing methods \citep{reverb_review}.

X-ray reverberation and the causal relationships between light curves $A$ and $B$ in two energy bands are measured via the \textit{cross spectrum} between those two light curves. Writing the Fourier transform of each light curve as the product of the amplitude and phase of each Fourier frequency component that makes up the stochastic variability in the light curve, $\tilde{A}(\omega) = |A(\omega)|e^{-i\varphi}$ and $\tilde{B}(\omega) = |B(\omega)|e^{-i\theta}$, the cross spectrum is written
\begin{align}
\tilde{C}(f) = \tilde{B}^*\tilde{A} = |\tilde{A}(\omega)||\tilde{B}(\omega)|e^{i(\theta - \varphi)}
\end{align}
The cross spectrum is binned by frequency. The phase lag between the two light curves at each frequency is given by the argument of the cross spectrum, which can be converted to a time lag as a function of frequnecy, the \textit{lag-frequency spectrum} by
\begin{align}
\tau(\nu) = \frac{\arg[\tilde{C}(\nu)]}{2\pi\nu}
\end{align}
The lag-frequency spectrum between two light curves can be interpreted as the time lag between correlated variability between the two light curves for the slow and fast components, or slow and fast variability processes, that make up the light curve. By convention, a positive time lag indicates that correlated variability in $A$ lags behind that in $B$ and $A$ is taken to represent the harder X-ray energy band.

X-ray reverberation from the inner regions of the accretion disc can be measured by computing the lag-frequency spectrum between light curves in energy bands dominated by reflection from the disc (either the 0.3-1\keV\ band dominated by a blend of relativistically broadened emission lines or the 4-7\keV\ band dominated by the relativistically broadened iron K fluorescence line) and by directly observed continuum emission (\textit{e.g} 1-4\keV).

The lag-frequency spectrum between two light curves with gaps can be estimated using Gaussian processes that have been independently fit to the two light curves. Once the hyperparameters have been optimised, a pair of continuous light curve samples are drawn; one from the Gaussian process for each energy band, and the lag-frequency spectrum is computed between them. The distribution of the time lag in each frequency bin is obtained by drawing successive pairs of samples from the Gaussian processes. The mean lag and associated confidence interval can be calculated in each bin.

In addition to the lag-frequency spectrum, the \textit{lag-energy} spectrum can be produced, showing the relative response times of successive energy bands to variability over a given frequency range. In this case, the cross spectrum is calculated between the light curve in each energy band and the light curve in a reference band and then averaged over the frequency range of interest before obtaining the time lag. To maximise the signal-to-noise, the reference band is typically taken to be the full energy range, though for each energy bin excluding that energy range from the reference band so as to avoid correlated errors.

The lag-energy spectrum can be estimated by fitting a Gaussian process to the light curve in each energy band. After fitting the hyperparameters, a continuous light curve is drawn from each Gaussian process, which can then be treated as if the observation had been continuous and evenly sampled. The reference band is constructed by summing the light curve samples at each energy, excluding the energy of interest, then the cross spectra and time lags are calculated. The lag-energy spectrum is sampled by continuously drawing sets of samples from the Gaussian processes fit to each energy band, allowing the mean lag and confidence interval to be calculated as a function of energy.

In addition to lag-frequency and lag-energy spectra, Gaussian processes fit to light curves in successive energy bands can be employed to obtain estimates of any spectral timing product calculated from the Fourier transforms of the light curves, including the covariance spectrum and the bispectrum following a similar procedure; draw sample light curves from the Gaussian processes then compute the product as if from continuous observations. While it is in principle straightforward to obtain the power spectrum or periodogram by a similar procedure, this is in practice unnecessary as this is given directly by the Fourier transform of the kernel function.

In practice there will be some uncertainty in the hyperparameters that are fit to what will be noisy data. This uncertainty can be included in the distribution of the lag spectrum by, instead of drawing light curve samples from Gaussian distributions with fixed hyperparameter values, drawing the hyperparameter values for each light curve draw from the posterior distribution for the hyperparameters. These posterior distributions are computed running a Markov Chain Monte Carlo (MCMC) sampler after the initial fit. The distribution of the lag spectrum will then be marginalised over the hyperparameters.

\subsection{Validation of Gaussian process method}
In order to validate the use of Gaussian processes to estimate the cross spectrum and measure X-ray reverberation from light curves with gaps, simulations were conducted based upon random light curves generated using the method of \citet{timmer_konig}. A random `continuum' light curve was produced with a broken power law PSD following $f^{-1}$ to a break frequency of $10^{-5}$\Hz, above which it falls off as $f^{-2}$. The generated time series was scaled to have the desired mean and standard deviation. The `reverberation' of this light curve was generated by convolving it with a $\delta$-function impulse response with the desired lag time. Gaps of regular length at regular intervals were introduced into both light curves by removing data points and each remaining data point was resampled from a Poisson distribution with mean corresponding to the simulated count rate to add the effect of Poisson noise.

Two types of AGN were simulated. The first, a typical, nearby narrow line Seyfert 1 galaxy in which X-ray reverberation from the inner accretion disc is frequently measured, based upon the AGN Ark\,564 \citep{kara+13}. Ark\,564 is observed by \textit{XMM-Newton} at a count rate of 39\ctsps\ with standard deviation of 12\ctsps\ over the 0.3-10\keV\ bandpass. For the purposes of these simple simulations, the count rate was split evenly between two light curves with 50\s\ binning, with a lag of 200\s\ between the light curves such that the lag spans multiple time bins.

Time lags in Ark\,564 are readily observable using the long, continuous light curve segnents available from \textit{XMM-Newton}, hence we here consider the measurement of time lags in this source by observatories in low-Earth orbit. We first consider the case of the \textit{Neutron Star Interior Composition Explorer} \citep[\textit{NICER},][]{nicer}. \textit{NICER} is a high throughput X-ray timing mission with peak effective area at 1\keV\ around 50 per cent larger than \textit{XMM-Newton}. While its primary mission is the study of the X-ray variability from neutron stars, it is highly suited to the detection of X-ray reverberation around black holes and has detected high frequency X-ray reverberation from the accretion disc in the stellar mass black hole transient MAXI\,J1820+070 \citep{kara_maxireverb}. 

Secondly, we consider the proposed specialised probe-class X-ray timing mission, \textit{STROBE-X} \citep{strobex}. Specifically designed for X-ray timing studies of black holes across the mass scale, neutron stars and transient events, \textit{STROBE-X} will possess two main instruments; a 3\msq\ \textit{X-ray Concentrator Array (XRCA)} sensitive between 0.2 and 10\keV\ and a 5\msq silicon drift detector, the \textit{Large Area Detector (LAD)}, sensitive between 2 and 30\keV. Due to its lower background and soft response, the \textit{XRCA} would be most suited to reverberation studies in AGN. \textit{STROBE-X} too would fly in low-Earth orbit, meaning that to fully leverage the capabilities of this large collecting area mission to study AGN, it will be necessary to combine timing data from multiple orbits.

Observations were simulated with both \textit{NICER} and the \textit{XRCA} on \textit{STROBE-X} by scaling the count rate seen in \textit{XMM-Newton} the effective area of these detectors. The count rate was estimated by fitting a simple model, comprised of a power law continuum component, reflection from the accretion disc and Galactic absorption, to the \textit{XMM-Newton} spectrum and folding the model through the appropriate response matrices using \textsc{xspec}. For \textit{NICER}, the count rate from Ark\,564 was simulated to be 65\ctsps\ and for \textit{XRCA} the count rate was 1230\ctsps. Regular gaps 2400\s\ in length were put into the light curve, beginning at 5700\s\ intervals to simulate low-Earth orbit.

The second class of AGN to be simulated is based upon the BLRG 3C\,120 with a count rate of 25\ctsps\ and standard deviation of 8\ctsps, observed over the 0.3-10\keV\ bandpass by \textit{XMM-Newton}. 3C\,120 has a black hole mass approximately 50 times that of Ark\,564, thus the lag between the two light curves was set at 5000\s\ for the test. The feasibility of conducting low frequency reverberation measurements by combining multiple successive orbits of \textit{XMM-Newton} observations  was simulated by inserting 35\ks\ gaps into the light curves every 172\ks.

\subsubsection{Reproducing the phase relationship}

If Gaussian processes are to successfully predict time lags that span gaps in light curves, the Gaussian process must maintain the phase relationship between the two light curves in its predictions.

The phase relationship between two light curves, $A(t)$ and $B(t)$, may be quantified through the \textit{coherence}, $\gamma^2$:
\begin{equation}
 \gamma^2 = \frac{|\langle \tilde{B}^*\tilde{A} \rangle |^2}{\langle|\tilde{A}|^2\rangle \langle |\tilde{B}|^2\rangle}
\end{equation}
The coherence is measured in finite-width frequency bins and angle brackets denote averaging over the frequency range in question. Once again, the tildes represent the Fourier transforms of the time series.

Coherence measures the fraction of the variability in one time series that can be predicted by applying a linear transformation (for instance convolving with a response function) to the other. If there is a consistent phase relationship between $A$ and $B$, the squared average over the cross spectrum in the numerator will have magnitude equal to $\langle|\tilde{A}|^2\rangle\langle|\tilde{B}|^2\rangle$ and thus the coherence will be unity. The coherence decreases to zero with more variability that is not produced by linear transformation of the other light curve (\textit{e.g.} with more uncorrelated noise).

If the Gaussian process accurately reproduces the phase information of the underlying continuous light curve, or the phase relationship between the two light curves from which lag measurements are to be made, the coherence between the two predicted light curves should be maintained at the level of the original, continuous curve.

\begin{figure*}
\centering
\subfigure[Ark\,564 \textit{NICER}] {
\includegraphics[width=85mm]{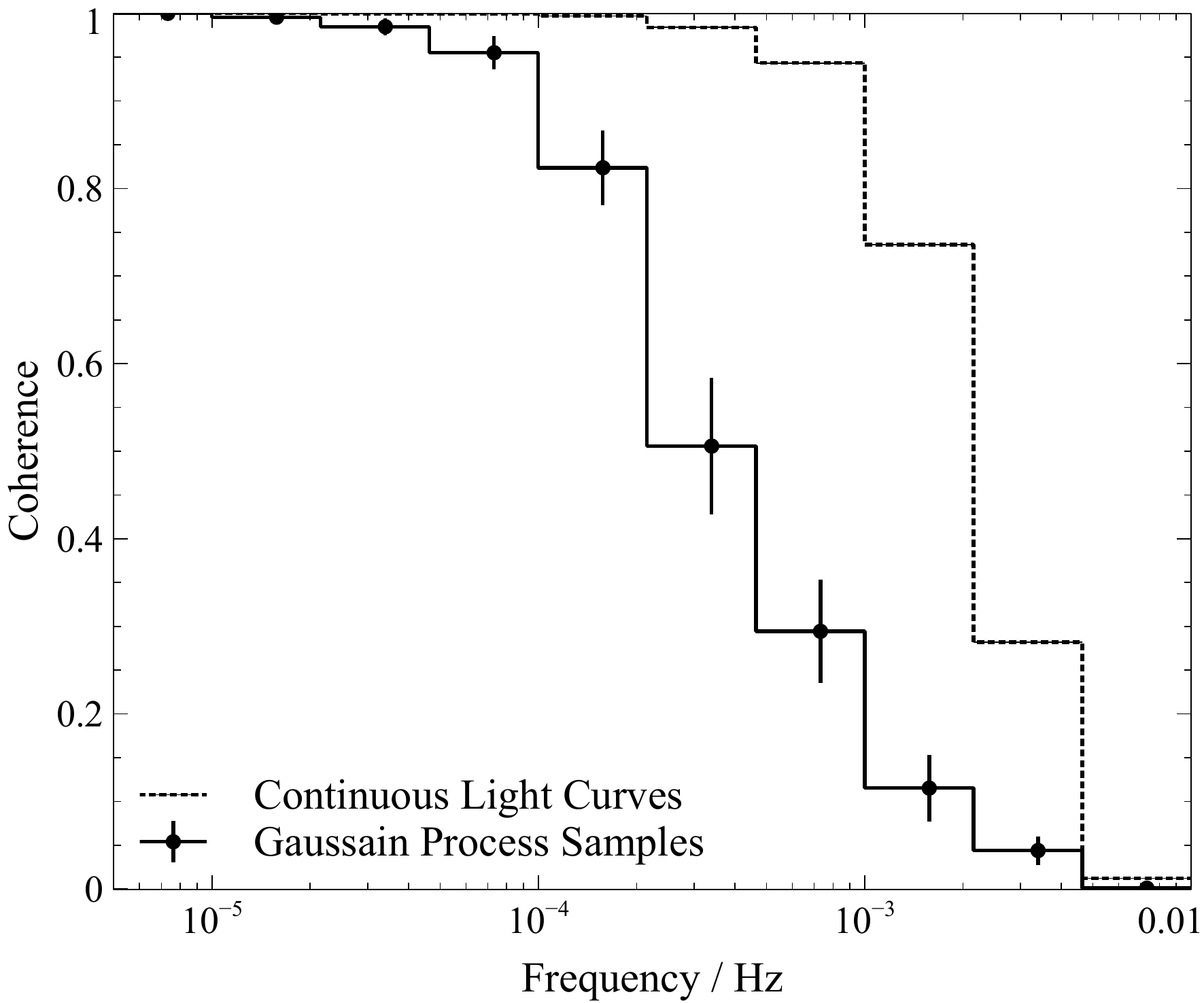}
\label{coherence_sim_ark564.fig:nicer}
}
\subfigure[Ark\,564 \textit{STROBE-X XRCA}] {
\includegraphics[width=85mm]{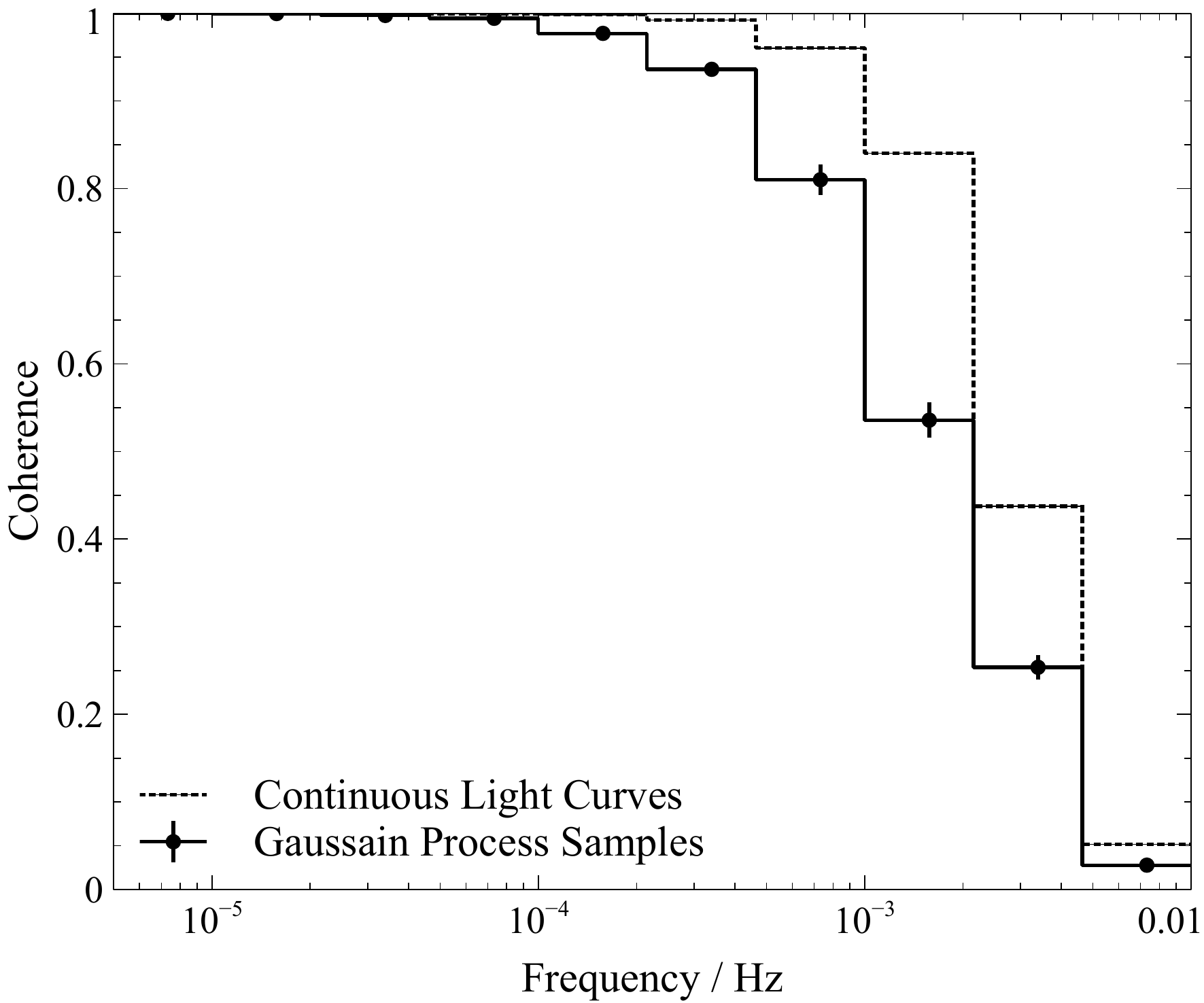}
\label{coherence_sim_ark564.fig:xrca}
}
\caption[]{The coherence, quantifying the phase relationship between pairs of sample light curves drawn from Gaussian processes, compared to the coherence between the original time series with no gaps for a simulated observation of an AGN like Ark\,564 with a 200\s\ lag between the light curves, observed with \subref{coherence_sim_ark564.fig:nicer} \textit{NICER} and \subref{coherence_sim_ark564.fig:xrca} the \textit{STROBE-X XRCA}. Error bars correspond to the standard deviation of the Gaussian process samples of the coherence.}
\label{coherence_sim_ark564.fig}
\end{figure*}

After Gaussian processes were fit to the pairs of simulated light curves, continuous sample light curves were drawn from the model distributions and the coherence between them was calculated as a function of Fourier frequency. The coherence between Gaussian process samples fit to the light curves of an AGN similar to Ark\,564 with a 200\s\ lag, observed by an instrument similar to \textit{NICER} in low-Earth orbit is shown in Fig.~\ref{coherence_sim_ark564.fig}.

Coherence is maintained close to unity between the pairs of sample light curves drawn from the Gaussian processes up to a frequency of $10^{-4}$\Hz, at which point it starts to drop. At higher frequencies, the characteristic timescale does not span a gap, meaning that the phase relationship between the two light curves is less constrained by the observed data points. The higher the frequency, the greater the contribution to the coherence from pairs of time bins lying only within a single gap, with which predictions will be less accurate than when timescales force the inclusion of a time bin in which a data point is observed.

\citet{epitopakis+2016} show that the phase lag between two light curves can be reliably measured if the coherence in a frequency bin is greater than $1.2/(1+0.2m)$ where $m$ is the number of Fourier frequencies averaged into the bin. 16 frequencies are included in the $(1-2)\times 10^{-4}$\Hz\ bin from a 100\ks\ light curve, suggesting that phase lags can be reliably measured if the coherence exceeds 0.3 (this value decreases for longer exposures). The coherence measured between the pairs of sample light curves drawn from the Gaussian processes suggest that phase estimates are reliable, given non-linear deviations introduced between the two light curves, up to a frequency of $10^{-3}$\Hz.

When observed with the large collecting area of the \textit{STROBE-X XRCA}, as shown in Fig.~\ref{coherence_sim_ark564.fig:xrca}, coherence values close to unity are maintained to higher frequencies, up to $4\times 10^{-4}$\Hz\ and once again maintaining values above 0.3 suggesting that reliable phase estimates can be obtained up to just over $10^{-3}$\Hz.

\begin{figure}
\centering
\includegraphics[width=85mm]{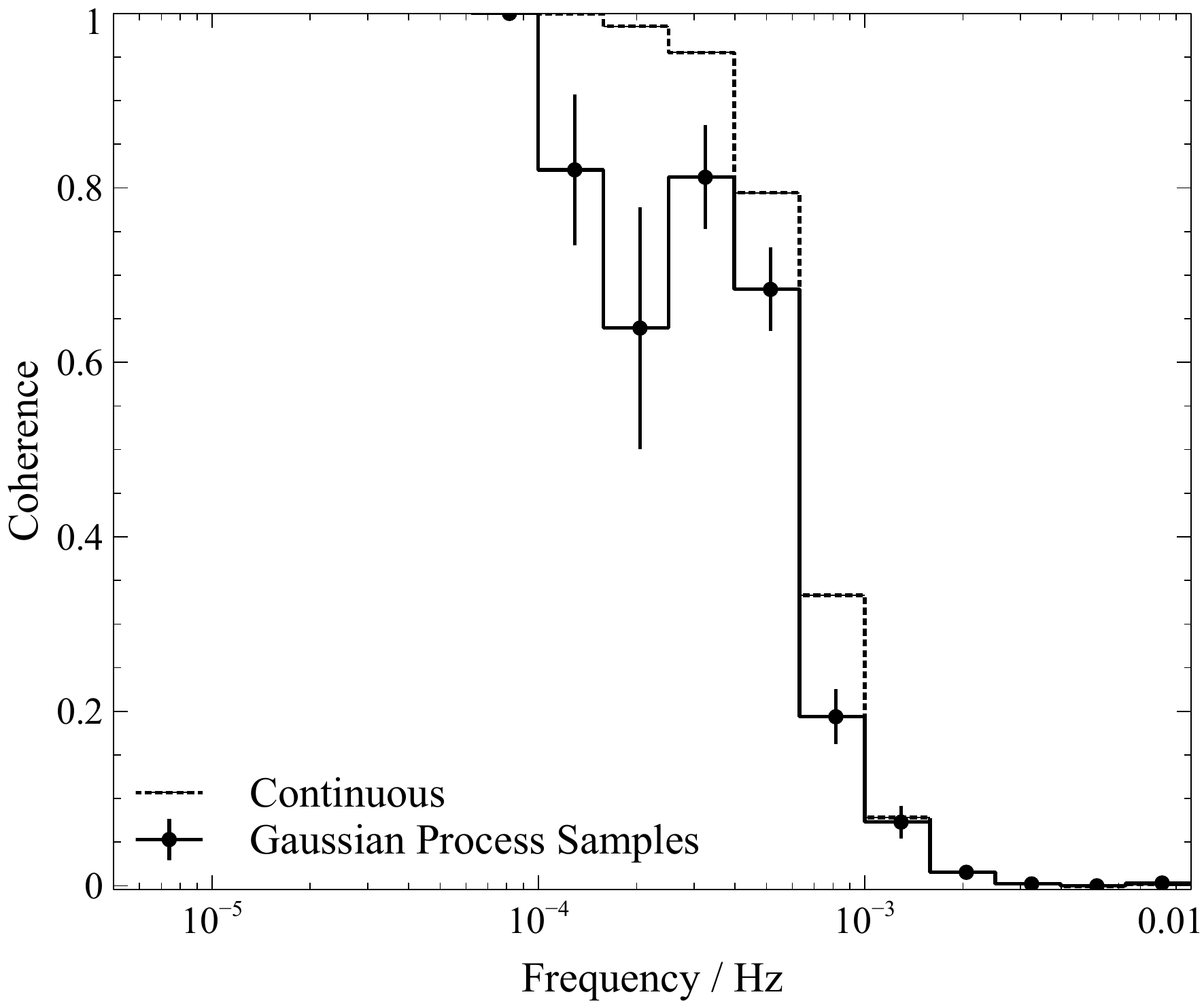}
\caption[]{As Fig.~\ref{coherence_sim_ark564.fig}. The coherence between pairs of simulated light curves for a simulated observation of 3C\,120 with a 5000\s\ lag between the light curves, observed with \textit{XMM-Newton} comparing samples drawn from the Gaussian processes to the original continuous time series.}
\label{coherence_sim_3c120.fig}
\end{figure}

Where longer, less frequent gaps are imparted in light curves with a long time lag of 5000\s\ by observation with a satellite such as \textit{XMM-Newton} (Fig.~\ref{coherence_sim_3c120.fig}), the longer continuous light curve segments means that the coherence is maintained at a higher level and is comparable to that which can be obtained with continuous light curves up to $10^{-3}$\Hz. At this point Poisson noise starts to dominate the cross spectrum and the coherence drops. Above $10^{-3}$\Hz, coherence is maintained at the required level to be able to obtain a reliable measurement of the phase lag between the two light curves. The coherence peaks at only around 0.8 as orbital gaps between segments limit the information available in the observed light curves about the lowest frequency Fourier components.

\subsubsection{Simulated lag measurement}
The lag-frequency spectrum was predicted from Gaussian processes fit to the simulated light curves. A Gaussian process was fit to both the original and lagged light curves. Continuous light curve samples (predictions) were drawn from each Gaussian process and for each pair of predictions, the lag-frequency spectrum was computed. The mean and standard deviation of the lag in each frequency bin was computed over the ensemble of predictions to produce the predicted lag-frequency spectrum and associated statistical uncertainty.

\begin{figure*}
\centering
\subfigure[Ark\,564, \textit{NICER}] {
\includegraphics[width=85mm]{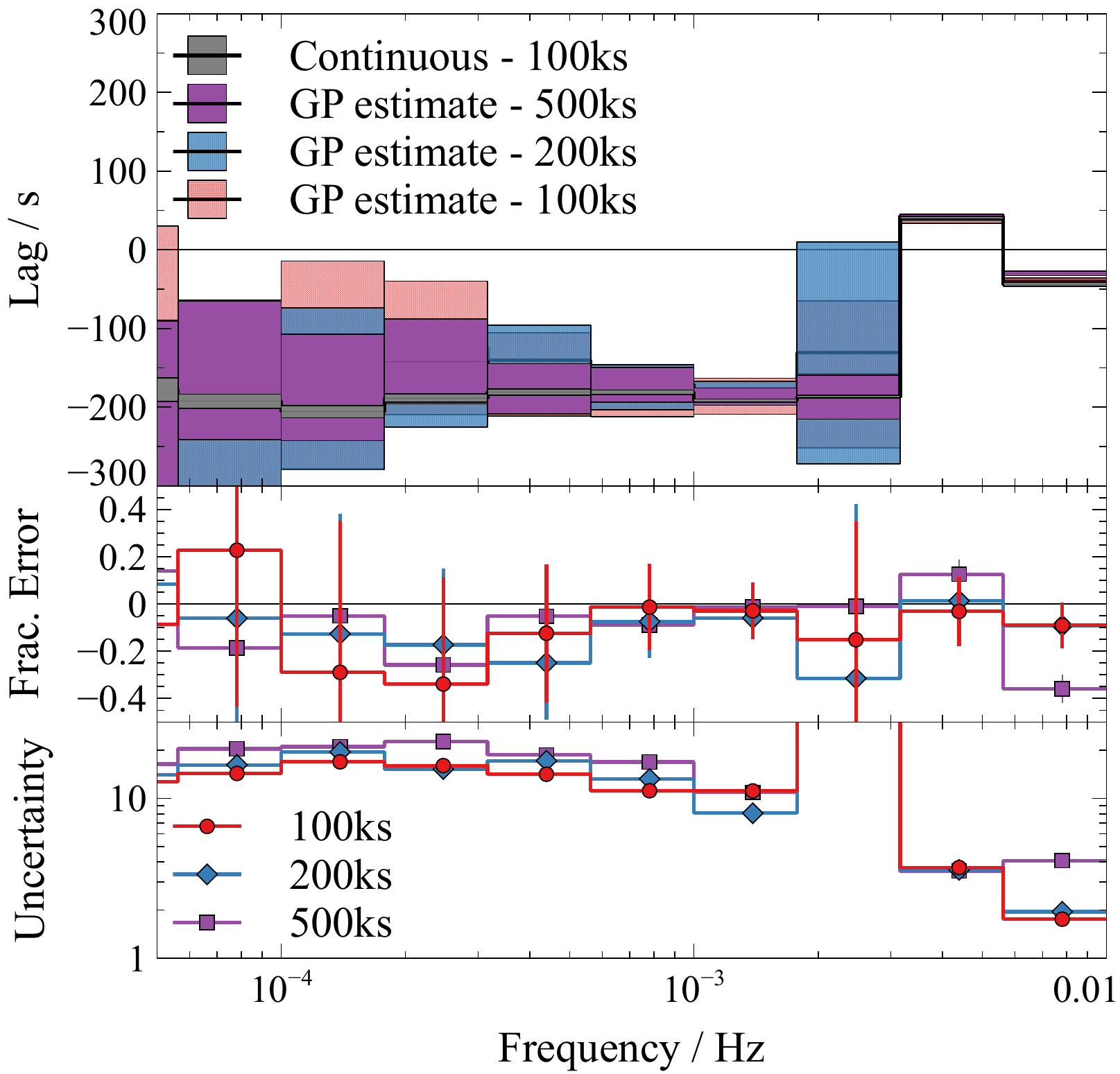}
\label{ark564_sim.fig:nicer}
}
\subfigure[Ark\,564, \textit{STROBE-X XRCA}] {
\includegraphics[width=85mm]{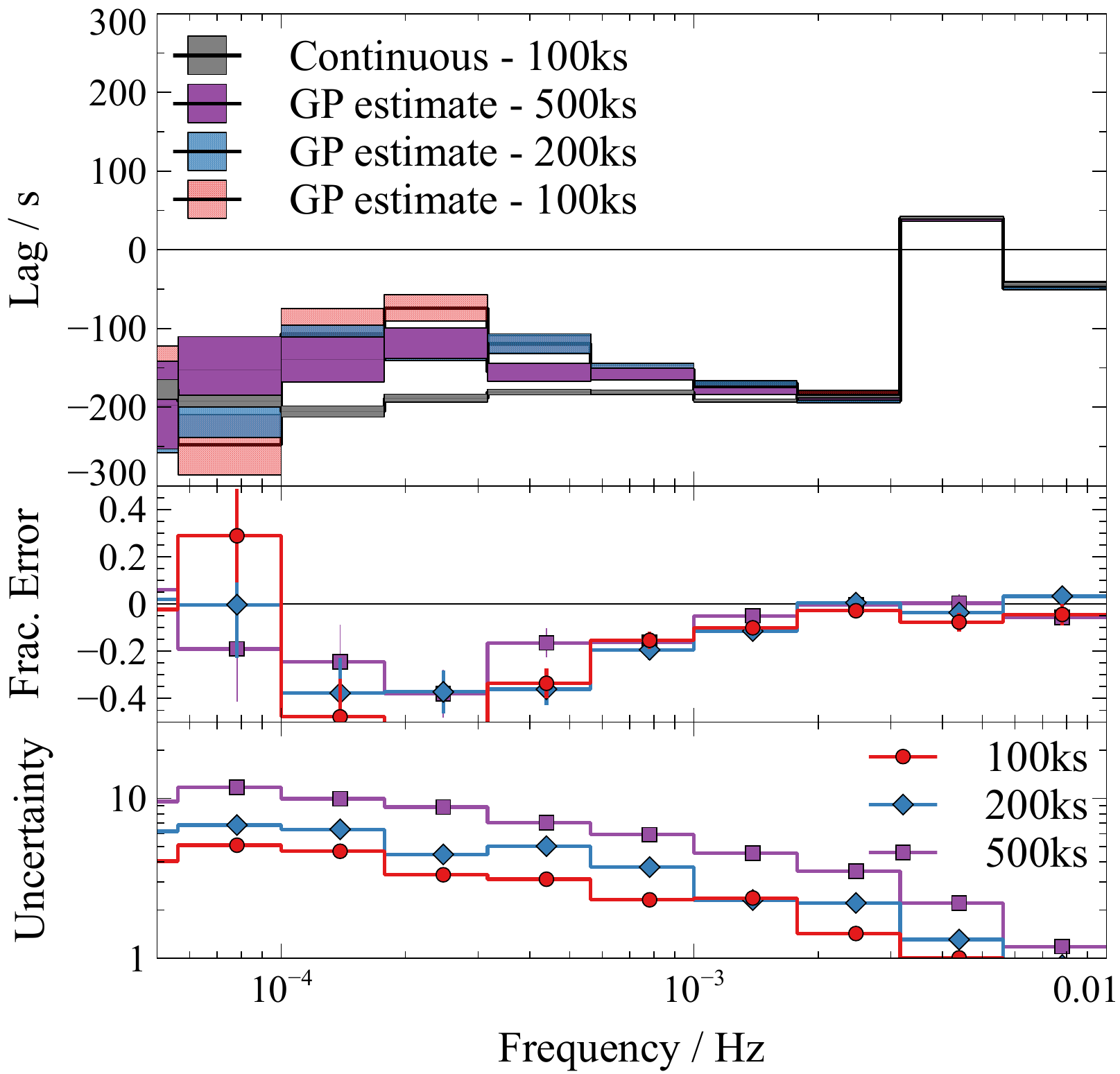}
\label{ark564_sim.fig:xrca}
}
\caption[]{The lag-frequency spectrum estimated by Gaussian processes fit to pairs of simulated light curves representing observations of the nearby, bright Seyfert galaxy Ark\,564 with a 200\s\ lag using \subref{ark564_sim.fig:nicer} \textit{NICER} and \subref{ark564_sim.fig:xrca} the large collecting area \textit{XRCA} on board the proposed \textit{STROBE-X} X-ray timing mission. In each case, the top panel shows the estimated lag as a function of frequency, along with the statistical uncertainty, obtained using 100, 200 and 500\ks\ of obervations, compared to a continuous observation lasting 100\ks. The middle panels show the fractional systematic error between the Gaussian process estimate and the lag computed from a continuous light curves with the same exposure. The bottom panels show the statistical uncertainty inferred from the Gaussian process sample as a multiple of the statistical uncertainty obtained using a single, continuous light curve with the equivalent exposure.}
\label{ark564_sim.fig}
\end{figure*}

The predicted lag-frequency spectra from the simulated light curves for an Ark\,564-like AGN with a 200\,s lag, observed with both \textit{NICER} and the \textit{STROBE-X XRCA} are shown in Fig.~\ref{ark564_sim.fig}. For each instrument, the top panel shows the lag-frequency spectrum that is estimated using 100, 200 and 500\ks\ total exposure (not counting the gap intervals), compared to the `correct' lag frequency spectrum that is computed applying the standard Fourier method to a continuous light curve with no gaps. The middle panel shows the fractional systematic error between the Gaussian process estimate and the true lag-frequency spectrum which for each exposure is calculated using the Fourier method from a continuous light curve with the same exposure. The bottom panel compares the statistical uncertainty (the computed size of the error bar) obtained from the standard deviation of the Gaussian process samples, as a multiple of the error bar calculated for the equivalent continuous light curve from the coherence \citep{nowak+99, reverb_review}.

For count rates obtained with an instrument akin to \textit{NICER}, it can be seen that the Gaussian process estimate of the time lag as a function of Fourier frequency generally agrees well with that computed from the equivalent continuous light curve. Systematic offsets from the true lag are below 20 per cent, except in the $2-3\times 10^{-4}$\Hz\ frequency bin where the lag is underestimated with a fractional error of 30 per cent, but below the level of the statistical uncertainty inferred from the standard deviation of the Gaussian process estimates.

The statistical uncertainty in the lag is typically about 10 times greater than it would be using continuous light curves. The gaps in the light curves add significant uncertainty, particularly at low frequencies. Low frequency components span across a gap in the light curve and hence there are significantly fewer pairs of data points at time separations that probe these Fourier frequencies than there would be were the gaps not present. At high frequencies, the uncertainty falls to only around 1.5 times that which could be achieved with a continuous light curve; at these frequencies, the corresponding timescale does not span a gap and the error is simply increased by approximately the square root of the effective exposure due to the missing data. The time lag is well measured between $5\times 10^{-4}$ and $2 \times 10^{-3}$\Hz, with fractional error less than 2 per cent, above which it is not possible to measure the 200\s\ lag due to phase wrapping.

When the count rate is increased to the level that would be expected from Ark\,564 if observed with the much larger collecting area of the \textit{STROBE-X XRCA}, the statistical uncertainty becomes much smaller (though still between three and 10 times that which would be achieved with a continuous light curve at this count rate at low frequencies, noting that for each exposure the uncertainty is compared to a continuous light curve with the same equivalent exposure, hence appears worse for longer exposures). The systematic error between $10^{-4}$ and $10^{-3}$\Hz\ becomes much more pronounced, peaking at 50 per cent for a 100\ks\ exposure, or around 30 per cent for longer exposures in the $2-3\times 10^{-4}$\Hz\ frequency bin. This systematic error appears at frequencies corresponding to the period between gaps in the low-Earth orbit light curve (approximately $1/5700$\Hz). The bias is negative, decreasing the measured lag value, since the gaps introduce a zero-lagging coherent signal between the light curves in the two energy bands. The bias, however, decreases to the 10-20 per cent level at both lower and higher frequencies and is less significant for longer light curve exposures. In this case, it appears that the error in the reproduction of the lightcurve autocorrelation by the Gaussian process is below the level mandated by statistical uncertainties in the data. This could likely be remedied by the selection of a kernel function that more accurately represents the data, albeit with less computational efficiency. Improvements to the kernel function for high signal-to-noise data will be explored in subsequent work. The accuracy that can be achieved using the rational quadratic kernel is sufficient for the analysis of data available from current missions.

\begin{figure}
\includegraphics[width=85mm]{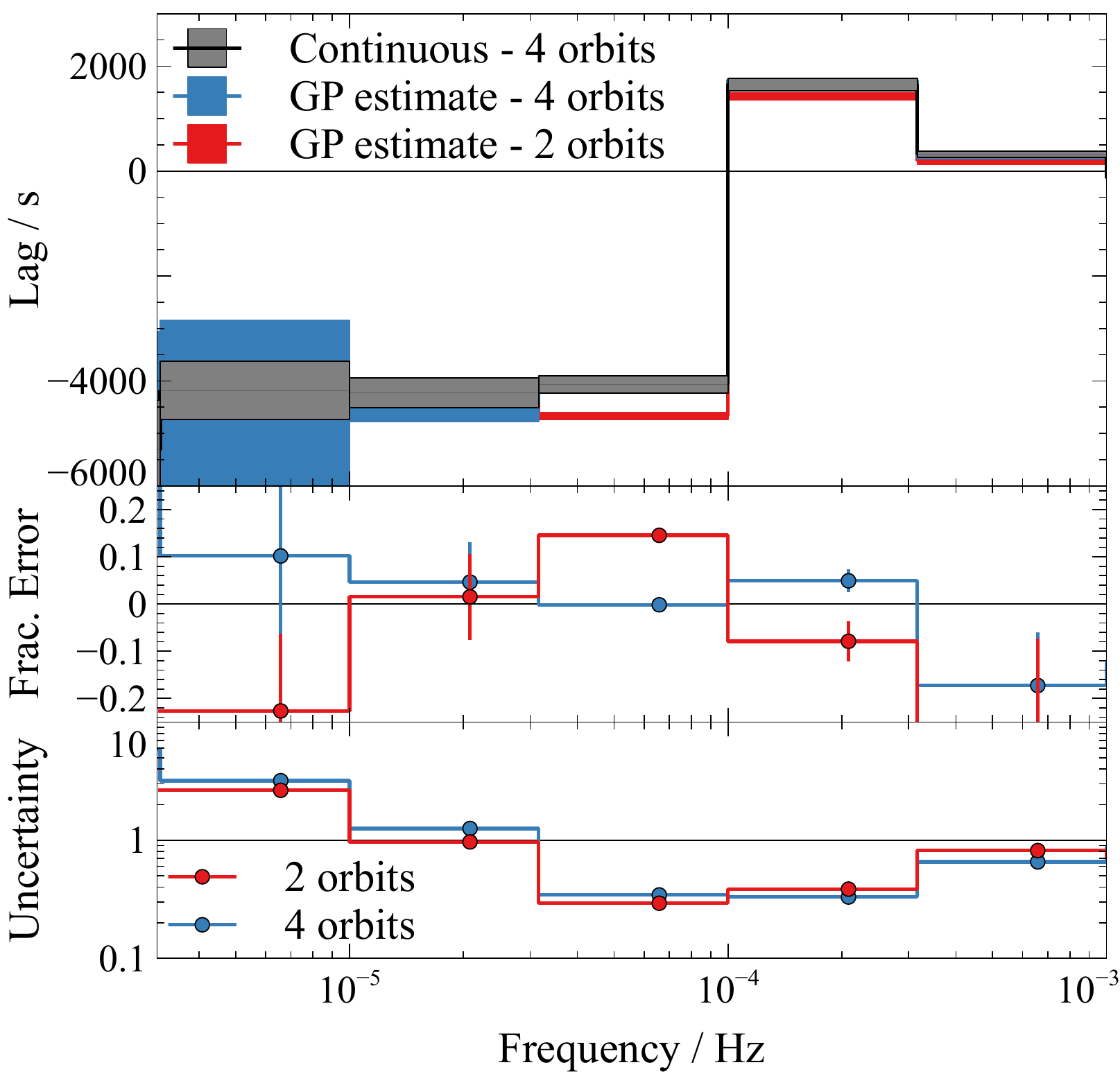}
\caption[]{As Fig.~\ref{ark564_sim.fig}, showing the lag-frequency spectrum estimated from Gaussian processes fit to a simulated pair of light curves with a 500\s\ lag and count rate equivalent to an obsrvation of the BLRG 3C\,120 by \textit{XMM-Newton}.}
\label{3c120_sim.fig}
\end{figure}

Turning to the case of long-timescale, low frequency X-ray reverberation that could be observed across multiple 130\ks\ orbits with \textit{XMM-Newton}, Fig.~\ref{3c120_sim.fig} shows the simulated measurement of the lag-frequency spectrum using Gaussian processes applied to light curves of an AGN like 3C\,120 with a 5000\s\ lag. 

The 5000\s\ lag is detected at frequenies below $10^{-4}$\Hz, above which phase-wrapping makes such a long lag undetectable. The lag is reproduced accurately, with systematic error less than 5 per cent above $10^{-5}$\Hz. Between $3\times 10^{-6}$ and $10^{-5}$\Hz, when four light curve segments are included in the analysis, the systematic error remains low at around 10 per cent. However, when only two segments are included (with just one gap), the systematic error increases to 20 per cent in this lowest frequency bin (and also increases to 10 per cent around $10^{-4}$\Hz). Errors at lower frequncies are greater when only two light curve segments are available, spanning just one gap. Few pairs of data points that span the gap with the appropriate time separation to probe these frequencies are available. In these observations, the bias that was seen in the low-Earth orbit case due to the orbital gaps is not expected until frequencies below $10^{-6}$\Hz, lower than the frequencies required to measure a 5000\s\ lag. 

The statistical uncertainty is around three times that expected from the equivalent continuous light curves with no gaps, though the statistical uncertainty estimated from the Gaussian process is about half that expected from the continuous light curve when the statistical uncertainty is calculated from the coherence using Equation 12 of \citet{reverb_review}.

To test this result, the error due to Poisson noise in the continuous light curves was recalculated by resampling each of the light curves. Each light curve sample was computed by drawing the count rate in each time bin from a Poisson distribution with mean equal to the original count rate (to simulate the scatter in each point due to noise, which is the only source of uncorrelated variability between the two simulated light curves). The lag-frequency spectrum was then calculated for each pair of resampled light curves and the $1\sigma$ uncertainty calculated from the sample. This test reveals that the coherence in fact overestimates the statistical uncertainty on such a long lag. Comparing the statistical uncertainty within the Gaussian process sample to that obtained by resampling continuous light curves, we find that the Gaussian process produces statistical uncertainty around 10 times that obtained with continuous light curves, in line with that seen where the gaps are shorter and more frequent.

Finally, a null test was performed to confirm the ability to reliably detect a lag between two light curves. Different Poisson noise realisations were created of the same underlying light curve (with gaps) and the lag-frequency spectrum was computed between them, to confirm the abililty to detect no lag. Zero lag was measured from these simulated light curves with no systematic offset and statistical uncertainty comparable to that obtained in the tests when a lag was present.

\subsection{The flux distribution}
The observed fluxes from accreting black holes are typically found to follow log-normal rather than normal distributions \citep{uttley+2005}. In order to ensure the accuracy of predictions made by Gaussian process models of light curves, it is important to verify that the flux distribution of the input data can be reproduced.

A Gaussian process with no constraint is expected to produce normally distributed time series since the multivariate Gaussian from which the sample light curves are drawn is formed from the union of the Gaussian distributions of the count rates in each time bin. In order to produce log-normally distributed light curves, the Gaussian process is fit to the logarithm of the observed light curve, then the sample light curves are produced by exponentiating the sample drawn from the Gaussian process.

In order to test the reproduction of the flux distribution of the time series, a Gaussian process was fit to a 200\ks\ simulated light curve with broken power law PSD and gaps corresponding to observation from low-Earth orbit, as above. A sample continuous light curve was drawn from the Gaussian process and a Kolmogorov-Smirnov (KS) test was applied to compare the count rate distribution of the sample light curve to that of the original light curve before the gaps were introduced.

When the input light curve possesses a normal count rate distribution, the KS statistic between the sample time series and the original was calculated to be 0.0146, representing a 91.8 per cent probability that the time series were drawn from the same underlying distribution.

When the input light curve possesses a log-normal count rate distribution, simulated by exponentiating the output of the Timmer \& Konig algorithm, the sample light curve still approximately reproduces the underlying flux distribution of the input. The KS statistic was found to be 0.0158, representing 86.8 per cent probability that the underlying distributions match. Interestingly, the Gaussian process does not produce a normal flux distribution and is able to approximately reproduce the correct distribution. The correct flux distribution is imparted by the observed data points in the conditional distribution from which the sample light curve is drawn (Equation~\ref{conditional.equ}).

When the Gaussian process is fit to the logarithm of the count rate, the agreement of the flux distribution improves with a KS statistic of 0.0143, corresponding to a probability of 93.3 that the sample and the original are drawn from the same distribution. In this instance, however, the random noise added to each time bin by the diagonal component of the kernel function will not possess the correct Poisson distribution. The Gaussian noise is now added to the logarithm of the count rate, not to the observed count rate. Thus, applying a Gaussian process to the logarithm of the count rate is only formally correct in the limit of high signal to noise (where the noise contribution is small). Experimentally, however, we find that this inaccuracy in the noise distribution does not introduce systematic errors into phase and time lag estimates that are obtained. It is sufficient to have enough freedom in the diagonal of the covariance matrix such that the Gaussian process can fit the underlying correlations between time bins without being forced to the incorrect form by uncorrelated noise.

Experimentally we find that either fitting the Gaussian process to the logarithm of the count rate (in which case any bins with zero detected counts are implicitly removed) or explicitly removing the time bins with zero count rate improves the statistical uncertainty in the phase measurements. When count rates are low, there is little information in the zero time bins since a photon count of zero could represent any rate below the detectable threshold. Not fitting the zero bins removes this uncertainty in the true count rate at these times and improves the phase constraint.

\section{X-ray reverberation in 3C\,120}

3C\,120, the bright, nearby, broad line radio galaxy is host to a supermassive black hole of mass $(5.7 \pm 2.7)\times 10^7$\Msun\ \citep{pozonunez+2012}. The iron K reverberation timescale seen from the accretion discs around supermassive black holes typically corresponds to the light travel time across between approximately 1 and 9\rg\ \citep{kara_global}. Hence in 3C\,120 the reverberation timescale expected to be as long as 3700\s. Such a long lag time would be detected in Fourier frequency components below $1/{2\tau}$, requiring accurate sampling of the light curve below frequencies of $1.4\times 10^{-4}$\Hz. It should be noted that the exact frequency range over which the reverberation lag from the accretion disc is detected depends upon the combination of processes driving the variability in the source in question and upon the exact form of the impulse response function that depends on the geometry of the corona and accretion flow \citep{cackett_ngc4151,propagating_lag_paper}. In 1H\,0707$-$495 ($M_\mathrm{BH} = 2\times 10^6$\Msun), the reverberation lag is most strongly detected at frequencies around $10^{-3}$\Hz\ \citep{fabian+09}. Scaling for the upper limit of the black hole mass suggests that a reverberation lag in 3C\,120 could be most strongly detected at frequencies as low as $2\times 10^{-5}$\Hz.

Such low frequencies are at the lower limit of what can be probed by stacking individual observation segments with \textit{XMM-Newton}, limited by the 130\ks\ orbit. Since few pairs of time bins are available to probe the lower limit of the frequency range in a single light curve segment, the uncertainty associated with lags measured at these frequencies is high. Gaussian processes were therefore employed to sample time lags in the low frequency Fourier components across multiple \textit{XMM-Newton} observations of 3C\,120.

\subsection{Observations}

\begin{figure*}
\centering
\includegraphics[width=175mm]{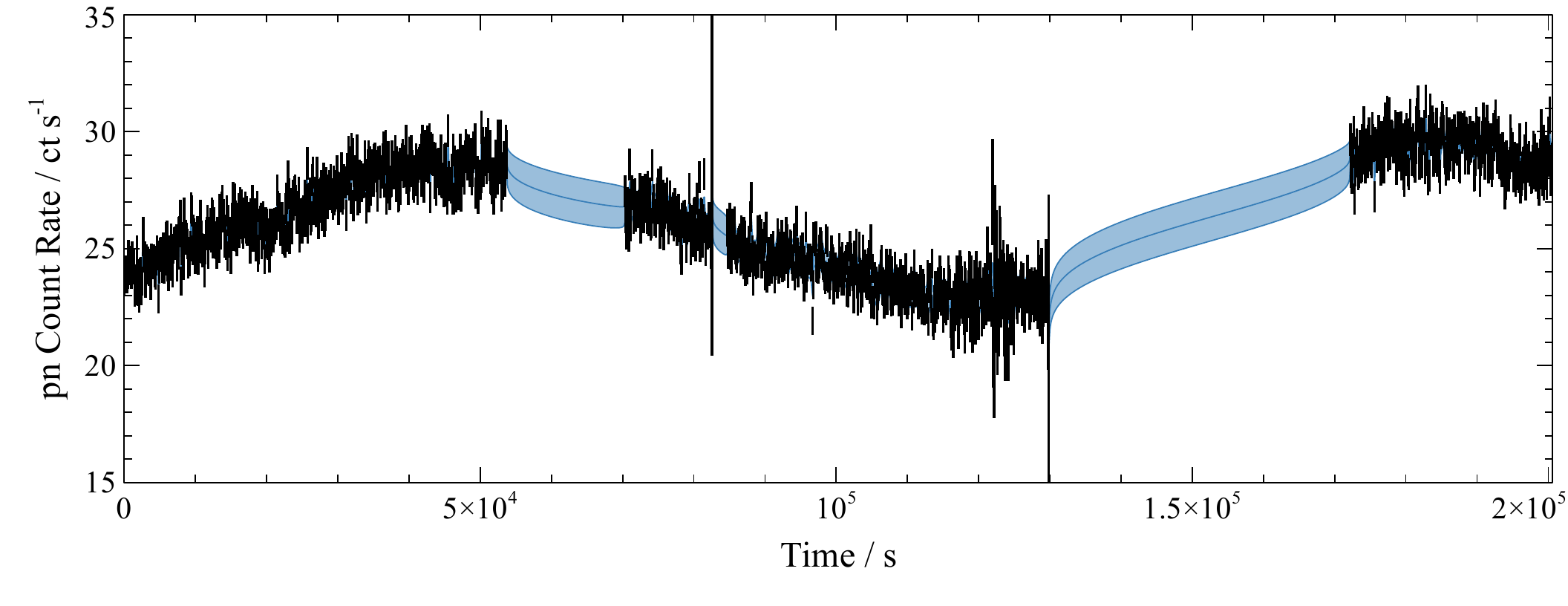}
\caption[]{X-ray light curve of the BLRG 3C\,120 observed with the EPIC pn camera on board \textit{XMM-Newton} in the 0.3-10\keV\ energy band in 2013 February. Shaded regions show the predictions marginalised across 1000 samples drawn from a Gaussian process, with a rational quadratic kernel fit to the observed data, along with the $1\sigma$ variation of the samples.}
\label{3c120_lc.fig}
\end{figure*}

3C\,120 has been observed on three occasions by \textit{XMM-Newton}, for 12\ks\ in 2002, for 133\ks\ in 2003 (one segment) and for a total of 159\ks\ in 2013. We here focus only on the 2013 observations, where 3C\,120 was observed in three segments with the EPIC pn camera \citep{xmm_strueder}, spanning a total time of 200\,ks, detailed in Table~\ref{3c120_obs.tab} and shown in Fig.~\ref{3c120_lc.fig}. The first OBSID is divided into two segments where the operating mode of the EPIC pn camera was changed and there is a 16\ks\ gap between them.

\begin{table*}
 \centering
 \caption{\textit{XMM-Newton EPIC pn} exposures during the 2013 observations of the BLRG 3C\,120 for which Gaussian processes are applied to the observed light curves to measure the time lags between X-ray energy bands. For each observation (OBSID), the pn exposures are shown along with the mean count rate and fractional variability ($F_\mathrm{var}$) in the 0.3-10\keV\ energy band.}
 \begin{tabular}{cccccc}
  \hline
  \textbf{OBSID} & \textbf{Date} & \textbf{Segment} & \textbf{Exposure} & \textbf{Count Rate} & $F_\mathrm{var}$ \\
  \hline
  0693781601 & 2013-02-06 & 1 & 53.7\ks & 26.8\ctsps & $0.057\pm 0.006$ \\
  & & 2 & 59.5\ks & 23.4\ctsps & $0.18 \pm 0.02$ \\
  0693782401 & 2013-02-08 & 1 & 28.4\ks & 29.2\ctsps & $0.012\pm 0.002$ \\
  \hline
 \end{tabular}
 \label{3c120_obs.tab}
\end{table*}

Only the 2013 observations contain sufficient sampling of the low frequency variability. Data from the \textit{XMM-Newton} EPIC pn are selected in preference to other detectors and missions due to the enhanced effective area of this instrument providing sensitivity to variability in the source luminosity across a broad energy range. The long continuous orbits of \textit{XMM-Newton} give good temporal sampling across a wide range in frequency.

3C\,120 is a relatively bright X-ray source, providing good signal to noise across the \textit{XMM-Newton} bandpass. Fractional variability, $F_\mathrm{var}$ \citep{edelson_fvar}, is relatively low, averaging $0.141\pm 0.009$ across the three segments of the 2013 observations, but as low as 6 or even 1 per cent during some of the segments.

\textit{XMM-Newton} data were reduced following the standard procedure using the \textit{XMM-Newton} \textsc{Science Analysis System (sas)} v16.1.0 using the most recent calibration available at the time of writing. The EPIC pn event lists for each observation segment were filtered to select single and double pixel (X-ray) events and time intervals during which the particle background count rate flared were removed using the standard criterion (when the total count rate between PI channels 10,000 and 12,000 exceeded 0.4\ctsps).

A circular source extraction region, 35\,arcsec in diameter, was defined centred on the point source and a background region of the same size was selected on the same chip as the source. Light curves were extracted in specific energy bands and the count rate in each time bin was corrected for exposure and dead time using the \textsc{sas} task \textsc{epiclccorr}. The light curves across the three observation segments were then concatenated into a single time series per energy band, leaving gaps where data are not available, either due to orbital gaps or due to background flares. Gaussian processes were fit to the logarithm of the count rate in these concatenated light curves. Shaded regions in Fig.~\ref{3c120_lc.fig} show how the Gaussian process is able to predict the light curve in the gaps, reproducing the variance of the observed data points and the trends across the gaps.

\subsection{Lag-frequency spectrum}
To search for signatures of X-ray reverberation from the accretion disc, the lag-frequency spectrum between the 1.2-4\keV\ energy band, expected to be dominated by directly-observed continuum emission, and the 4-7\keV\ band where the relativistically broadened iron K$\alpha$ line produced by the irradiated inner accretion disc is seen \citep[see models of the X-ray spectrum in 3C\,120 in][]{lohfink_3c120} was calculated.

The light curve in each band was fit with a Gaussian process. A rational quadratic (RQ) kernel function, which was found to be most suitable for a realistic broken power law power spectral density where gaps in the observed light curve are longer than $5000$\s, was used. The lag-frequency spectrum was estimated from 5000 light curve samples drawn from the pair of Gaussian processes, each time selecting values of the hyperparameters from their posterior distributions. In computing the lag-frequency spectrum, the cross spectrum was averaged into seven logarithmically spaced frequency bins between $5\times 10^{-6}$ and $10^{-2}$\Hz.

\begin{figure}
\centering
\includegraphics[width=85mm]{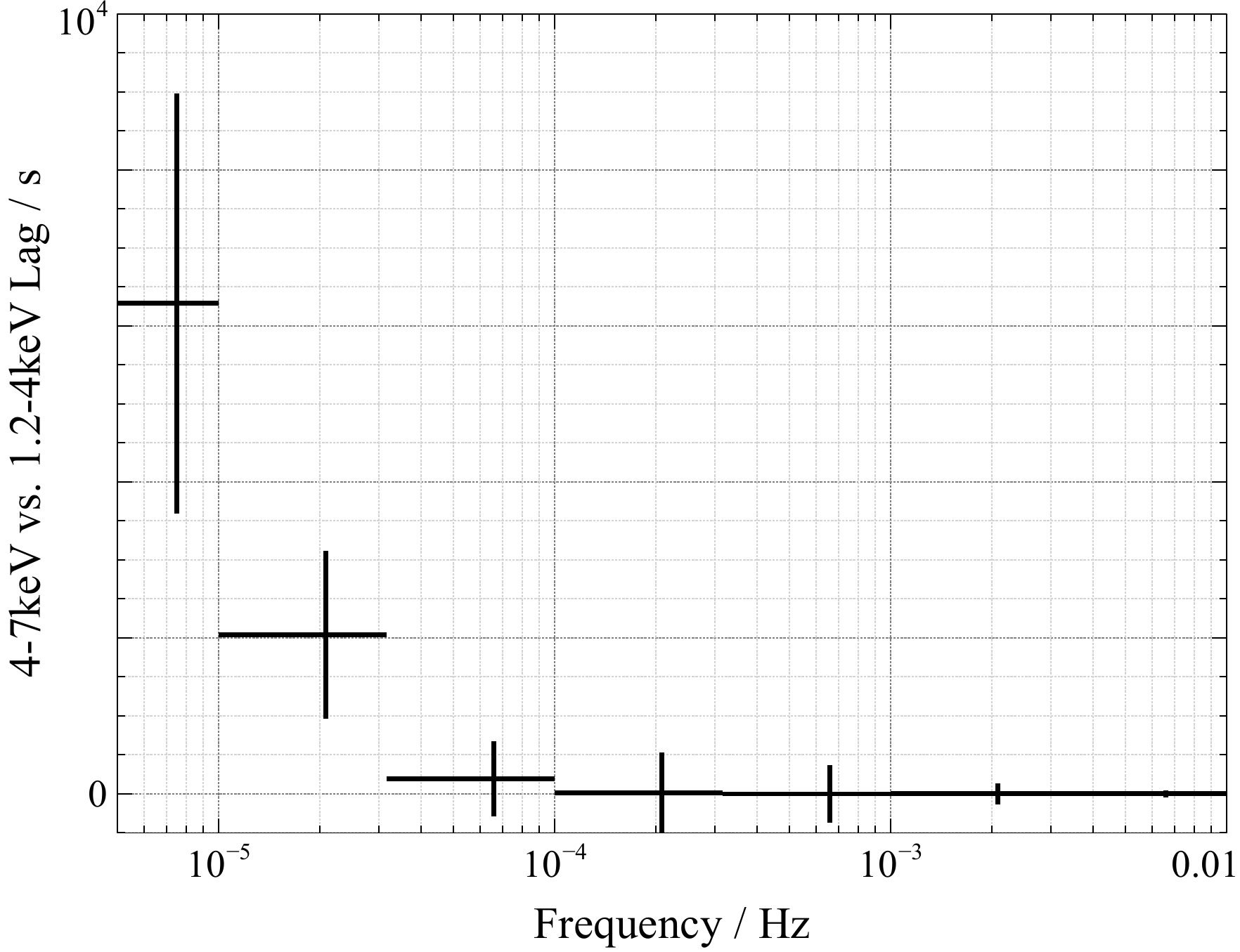}
\caption[]{The lag-frequency spectrum of 3C\,120 between the 1.2-4\keV\ energy band, dominated by the X-ray continuum, and 4-7\keV\ energy band dominated by the iron K line from the accretion disc, estimated from sample light curves drawn from Gaussian processes fit to the observed light curve segments.}
\label{3c120_lf.fig}
\end{figure}

Fig.~\ref{3c120_lf.fig} shows the lag-frequency spectrum between the 1.2-4\keV\ energy band and 4-7\keV\ energy band, estimated from Gaussian processes fit to the 2013 observations. A positive lag indicates that variations in the harder 4-7\keV\ band lag behind the correlated variations in the softer 1.2-4\keV\ band.

The time lag is found to increase from zero above $10^{-4}$\Hz\ to $(6300\pm2700)$\s\ at $7.5\times 10^{-6}$\Hz\ showing that at these low frequencies, variability in the 4-7\keV\ energy band lags behind that in the 1.2-4\keV\ band. Such a lag could be due to the reverberation of X-rays in the iron K fluorescent line and correspond to the additional light travel time from the primary X-ray source to the disc. Such a lag between these two energy bands could also, however, be due to the `hard lag' that is seen in the X-ray continuum emission in both AGN and X-ray binaries where higher and higher X-ray energies are seen to respond later and later to variations in luminosity and are likely due to the propagation of fluctuations through the corona.

\begin{figure}
\centering
\includegraphics[width=85mm]{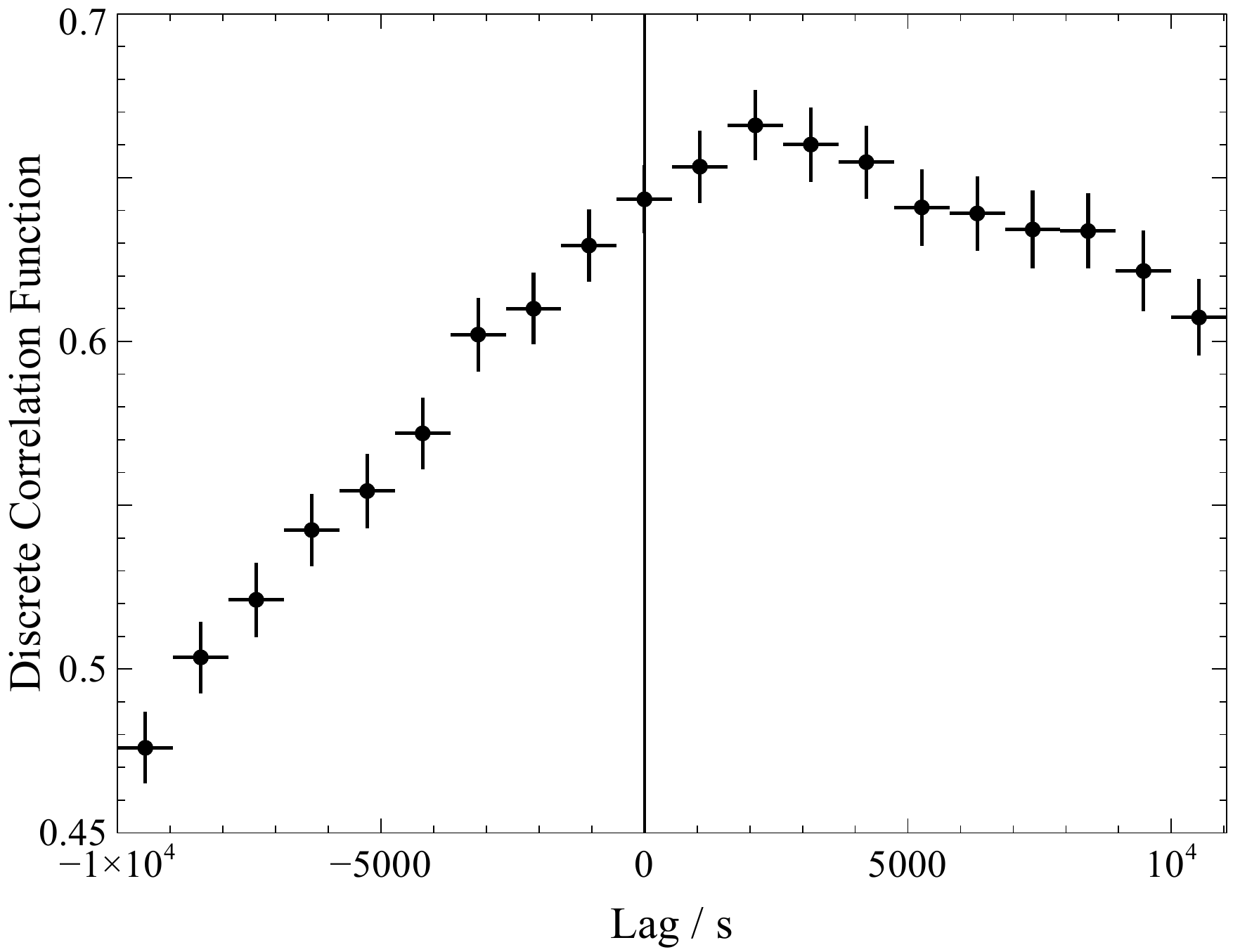}
\caption[]{The discrete correlation function (DCF) between the observed light curves in the 1.2-4\keV\ and 4-7\keV\ energy bands. Positive lags indicate variability in the hard band lagging behind that in the soft.}
\label{3c120_dcf.fig}
\end{figure}

In order to confirm this finding of a lag between variability in the 1.2-4\keV\ energy band and that in the 4-7\keV\ band, the time domain cross-correlation function can be examined. Since the light curves spanning the three observation segments are not evenly sampled, the discrete correlation function \citep[DCF, ][]{edelson_dcf} was computed, as shown in Fig.~\ref{3c120_dcf.fig}. Since the measured lag extends to low frequencies and has the same sign at all frequencies, it can be seen as a shift in the peak of the DCF from zero to a lag of $\sim 2500$\s, declining to a plateau around 7000\s.

\subsection{Lag-energy spectrum}
In order to determine the origin of the lag seen between the the 4-7\keV\ and 1.2-4\keV\ energy bands between $5\times 10^{-6}$ and $3\times 10^{-5}$\Hz, the \textit{lag-energy spectrum}, showing the relative response times of different energy bands to variability within this frequency range, was estimated from light curves extracted in 14 approximately logarithmically spaced energy bins between 0.3 and 10\keV. The lag-energy spectrum shows the average response time of variations in each energy band relative to the correlated variations in some reference band (and hence the zero point is arbitrary; it is the relative response times of successive energy bands that matters).

For each energy band, the reference band was taken to be the full 0.3-10\keV\ band, minus the energy band of interest so as to avoid correlated noise between the two bands as is common practice \citep{zoghbi+2012,kara+13,reverb_review}.

A Gaussian process was fit independely to the light curve in each energy band. 4000 continuous light curve samples were then drawn from the Gaussian processes and the lag for each energy bin was calculated following the standard procedure \citep{reverb_review}. The reference band was constructed from the sample and the average cross spectrum was computed across the frequency bin of interest. The estimated lag for each energy bin was taken to be the mean of the predictions across all of the samples and the negative and positive error bars were taken from the 15.9 and 84.2 percentiles for the $1\sigma$ spread of the distribution.

\begin{figure}
\centering
\includegraphics[width=85mm]{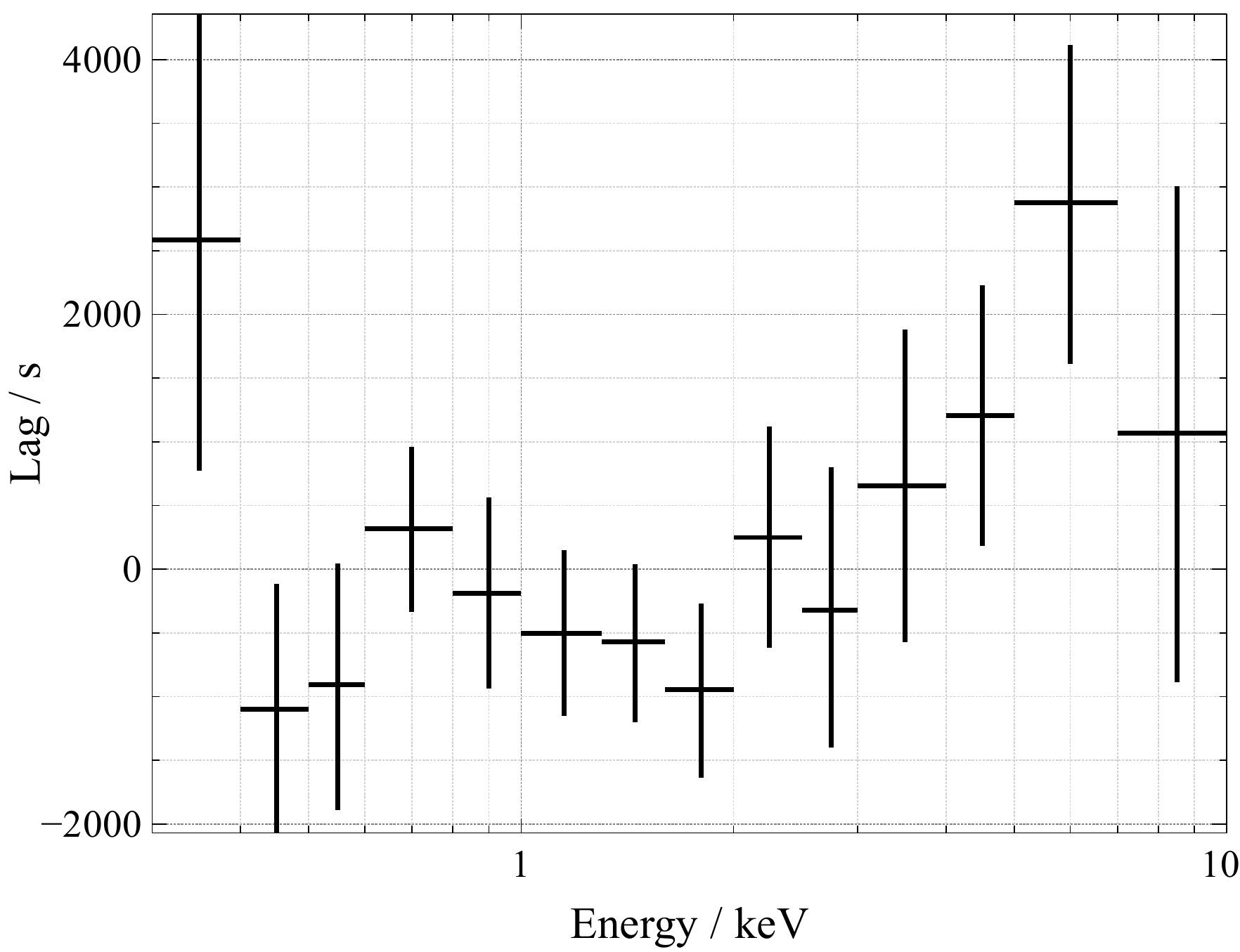}
\caption[]{The lag-energy spectrum of 3C\,120 over the $(5-30)\times 10^{-6}$\Hz\ frequency range (where the iron K band is seen to lag behind the energy band dominated by the X-ray continuum), estimated from Gaussian processes fit to the observed light curve segments in approximately logarithmically-spaced energy bins.}
\label{3c120_en.fig}
\end{figure}

The estimated lag-energy spectrum over the $5\times 10^{-6}$ to $3\times 10^{-5}$\Hz\ frequency band is shown in Fig.~\ref{3c120_en.fig}. The profile of the reverberation response from the accretion disc can be seen, following the form commonly seen in Seyfert galaxies \citep{zoghbi+2012,kara+13,kara_global}. 

The earliest response is seen in the continuum-dominated 1-2\keV\ energy band and the profile of the iron K line is apparent. The core of the line at 6.4\keV\ originates from the outer regions of the accretion disc, furthest from the illuminating X-ray source, so responds the latest to variations in luminosity. The redshifted wing of the iron K line, between 3 and 5\keV, is produced from the inner regions of the disc, closer to the X-ray source, hence responds earlier than the core of the line, but later than the directly-observed continuum between 1 and 2\keV. The line profile is seen as the response time of the 7-10\keV\ band drops relative to the 5-7\keV\ band. In addition there is some evidence for a delayed response from the soft excess, emitted from the accretion disc, with the response time increasing again at soft X-ray energies below 1.6\keV, though the response time drops again at 0.7\keV. The reverberating soft excess is expected to extend below this energy.

We can conclude that the reverberation response from the iron K line, emitted from the accretion disc illuminated by an X-ray emitting corona, is detected in 3C\,120 with a lag time of $(3800\pm 1500)$\s\ between the 5-7\keV\ and 1.6-2\keV\ bands. A hard lag within the continuum emission itself would have produced a lag-energy spectrum smoothly increasing with energy which is not seen here.

The significance of the detection of the iron K reverberation lag can be estimated by enumerating the samples in which the 5-7\keV\ energy bin seen seen to respond later than the 1.6-2\keV\ energy bin. Since sample lag-energy spectra are computed from each set of sample light curves, if the iron K reverberation feature is significant, the 5-7\keV\ bin will respond later in the majority of the sample spectra. The iron K lag feature was detected in 99.5 per cent of the sample indicating a significant detection.

The detection of the 5-7\keV\ iron K emission from the outer disc responding later than the 3-4\keV\ redshifted wing of the line is less significant, at the 88.9 per cent confidence level, while the 3-4\keV\ bin lags behind the 1.6-2\keV\ continuum-dominated bin at the 86.3 per cent confidence level. The redshifted iron K emission in the 4-5\keV\ bin lags behind the 1.6-2\keV\ continuum bin at the 96 per cent confidence level.

\section{Discussion}
Gaussian processes have been successfully employed to conduct X-ray timing analyses at frequencies below the limit imposed by the longest continuous light curve segments that are available due to orbital gaps in observations. A Gaussian process model is constructed, fitting a \textit{kernel function} to the observed data points that describes the autocorrelation or covariance matrix between pairs of time bins. Samples are then drawn from this multivariate probability distribution to obtain an estimate of the underlying continuous time series. Multiple, continuous, sample light curves can be drawn and timing analysis techniques based on the Fourier transforms of these continuous time series can be applied to each sample to obtain the distributions of the spectral-timing products. Specifically, the lag-frequency spectrum between a pair of light curves (the time lag as a function of the different frequency Fourier components that make up the light curves) was estimated by fitting Gaussian process models to the two light curves independently, then drawing continuous samples from each from which the lags were calculated.

In contrast to the method of \citet{zoghbi_gap} for estmating time lags between pairs of unevenly sampled light curves, the Gaussian process framework treats each light curve independently. Rather than fitting a model to the cross-correlation of the two light curves that parameterises the lag, the Gaussian process fits each light curve independently, with the only model assumption being the form of the kernel function or autocorrelation of the single time series. Once continuous light curve samples have been drawn, any of the standard timing analysis techniques based on their Fourier transforms can be applied, while uncertainties can be estimated by marginalising over multiple light curve samples drawn from the Gaussian processes.

Gaussian processes were applied to the energy-resolved light curves of the broad line radio galaxy (BLRG) 3C\,120 in order to measure the time lags between different X-ray energy bands. Variability in the iron K band was found to lag behind that in the continuum band at frequencies below $3\times 10^{-5}$\Hz.

The lag-energy spectrum shows the characteristic profile of X-ray reverberation from the inner regions of the accretion disc; most notably the characteristic shape of the relativistically broadened iron K line, formed by the combination of Doppler shifts and gravitational redshift in close proximity to the black hole. The latest response is seen around the rest frame energy of the iron K$\alpha$ line in the 5-7\keV\ energy band, lagging behind the continuum-dominated 1.6-2\keV\ band by $(3800\pm 1500)$\s. Line emission from the inner parts of the disc is more strongly redshifted and these regions are closer to the primary X-ray source, hence a shorter lag is seen in the redshifted wing of the emission line. 

A significant soft lag is not seen in the lag-energy spectrum of 3C\,120. In many other AGN, a time delay with respect to the continuum comparable to that in the iron K band is seen at X-ray energies below 1\keV\ \citep{kara+13}. This is attributed to the combination of relativistically broadened soft X-ray lines also reverberating from the disc. In 3C\,120, modelling of the X-ray spectrum suggests that the soft X-ray emission contains a significant contribution from synchrotron emission from the jet \citep{lohfink_3c120}. The comparable response time (\textit{i.e.} short time lag) between of the soft X-ray band and the coronal X-ray continuum suggests that these two emission components vary together and the coronal X-ray emission and jet emission are closely related.

A time lag of $(3800\pm 1500)$\s\ between the peak of the iron K line and the continuum, and a black hole mass of $(5.7 \pm 2.7)\times 10^7$\Msun, place the corona above the disc at a height of $(13\pm8)$\rg. Comparing this lag time to the sample of iron K lags in Seyfert galaxies \citep{kara_global} shows that the reverberation timescale and hence the characteristic scale-height of the corona above the disc \citep{lag_spectra_paper} is comparable between the BLRG 3C\,120 and radio quiet Seyfert galaxies. The lag in 3C\,120 and hence the scale-height of the corona, however, lies on the upper limit of the sample in which most lags are seen to lie below the light travel time over 9\rg. If the X-ray source is associated with the large-scale jet, one might speculate that the source of irradiation of the disc is higher than in a radio quiet AGN where there is no jet. For example, energy in the jet may be almost entirely in electromagnetic fields until a shock or instability develops some distance from the black hole, leading to its dissipation \citep[\textit{e.g.}][]{polko+2010}. This estimate of the scale-height does not account for dilution of the intrinsic lag by the contribution of continuum emission to the iron K band and \textit{vice versa} and therefore represents a lower limit to the coronal height.

In many radio quiet Seyfert galaxies, the earliest response time in the lag-energy spectrum is seen in the 3-4\keV\ energy band dominated by the redshifted wing of the iron K line, rather than in the 1-2\keV\ band that is most strongly dominated by the continuum \citep{kara+13}. \citet{propagating_lag_paper} show that this behaviour can be explained if the high frequency variability originates within a collimated core of the corona through which fluctuations in luminosity propagate upwards sufficiently slowly ($\sim 0.1c$). At lower frequencies, the variability in the continuum emission originates in an extended component of the corona above the inner parts of the accretion disc (out to $\sim 10$\rg) that is invoked to explain the illumination pattern of the disc required to produce the observed time-average iron K$\alpha$ line profiles \citep{1h0707_emis_paper,understanding_emis_paper} as well as the low frequency propagation lags in the continuum emission \citep{arevalo+2006,walton_hardlag}. Such an extended corona produces the earliest response time in the 1-2\keV\ band of the lag-energy spectrum.

In the radio-quiet Seyfert galaxy I\,Zw\,1, a transition is seen in the lag-energy profile as a function of frequency suggesting that the high frequency variability is dominated by the core of the corona while there exists simultaneously an extended component of the corona varying more slowly representing dissipation in the magnetosphere above the disc \citep{1zw1_corona_paper}. The compact, collimated core within the corona of radio quiet AGN could represent the dissipation of energy within the magnetosphere when a jet fails to launch due to the magnetic field configuration \citep{yuan_lp1, yuan_lp2}. 

Understanding the structure of the corona in a radio-loud AGN through X-ray reverberation will provide important constraints on such a model. In particular, being able to determine if the 3-4\keV\ redshifted wing of the iron K$\alpha$ line from the innermost parts of the disc leads the variability in the 1-2\keV\ continuum band will show if dissipation of energy in collimated X-ray emitting corona close to the black hole can exist alongside a jet or if the two are mutually exclusive.

X-ray emission could come from within the jet base. If fluctuations propagate up the jet sufficiently slowly, as in the case of radio quiet AGN, the earliest response will be detected in the 3-4\keV\ band. If the fluctuations propagate too rapidly, for example at the Alfv\'{e}n speed, which is likely close to the speed of light in a region dominated by magnetic pressure, the earliest response will be seen in the 1-2\keV\ band. 

If instead the only component of the corona is extended over the inner parts of the accretion disc, the earliest response will again be seen in the 1-2\keV\ band. Such a corona could be associated with reconnection in magnetic field lines emanating from the disc. It is likely that in order to support and collimate a large scale jet, significant vertical field from the disc is required (Blandford, private comm.), explaining the larger scale height of the corona above the disc than in radio quiet Seyfert galaxies.

In 3C\,120, there are suggestions that the earliest response is seen in the 1.6-2\keV\ energy band, rather than in the 3-4\keV\ band (as in the case of radio quiet Seyfert galaxies). This is only detected at 86.3 per cent confidence, however, so meaningful conclusions cannot be drawn, emphasising the need for further observations to provide a higher signal-to-noise estimate of the lag-energy spectrum in order to constrain the interplay between the jet and corona.

\section{Conclusions}

A framework has been developed employing Gaussian processes to perform Fourier-domain timing analyses on X-ray light curves with orbital gaps. The framework was used to measure time lags associated with X-ray reverberation from the accretion disc that are longer and are seen at lower frequencies that can be measured employing standard Fourier transform techniques to the longest continuous segments that are available. Gaussian processes naturally provide a probabilistic framework to obtain the posterior distributions of the time lags with which the significance and statistical uncertainties of results can be readily assessed.

The accuracy with which Gaussian process models can reproduce the phase relationship and time lag between simulated pars of light curves was verified. With the appropriate selection of kernel function for the red noise power spectra commonly observed in the X-ray variability of accreting black holes, namely the \textit{rational quadratic} kernel, it is possible to measure the time lag to within 2 per cent accuracy at frequencies between $5 \times 10^{-4}$ and $2\times 10^{-3}$\Hz\ in a typical bright AGN observed with an instrument such as \textit{NICER} in low-Earth orbit, introducing gaps into the light curve every 90 minutes. Gaussian processes also enable long time lags to be measured in high mass AGN from multiple, consecutive 130\ks\ orbits with \textit{XMM-Newton} to within 5 per cent accuracy at frequencies as low as $3\times 10^{-6}$\Hz.

Time lags between successive X-ray energies in the low frequency variability components were measured in the broad line radio galaxy 3C\,120. At frequencies between $5\times 10^{-6}$ and $3\times 10^{-5}$\Hz, the profile of the lag-energy spectrum reveals the reverberation of X-rays from the inner accretion disc. Reverberation in the relativistically broadened iron K line is significantly detected at the 99.5 per cent confidence level, with the 5-7\keV\ peak of the line lagging behind the 1.6-2\keV\ continuum-dominated energy band by $(3800\pm1500)$\s. Shorter time lags are seen in the redshifted wing of the line that originates from the inner accretion disc, closer to the primary X-ray source. The height of the primary X-ray source in 3C\,120 is estimated to be $(13\pm8)$\rg\ above the accretion disc.

The capability to probe X-ray reverberation at low frequencies and on long timescales that is afforded by Gaussian processes enables the study of X-ray reverberation to be extended from relatively low mass supermassive black holes in Seyfert galaxies to more massive black holes in radio galaxies and potentially quasars. Using X-ray reverberation to understand the differences in the location and structure of the X-ray emitting corona between radio-loud and radio-quiet AGN will yield important insight into the differences in the environments between these black holes and the mechanisms by which some black holes are able to launch jets.

The ability to conduct X-ray timing analyses on timescales that span multiple orbits will become increasing important with next-generation X-ray timing experiments in low-Earth orbit. \textit{NICER} is a high throughput X-ray timing experiment on the International Space Station which while primarily intended to study neutron stars, is ideally suited to conducting timing experiments on AGN. Great advances are promised by the proposed X-ray timing mission \textit{STROBE-X}, offering an order of magnitude enhancement in effective area at 6\keV\
over even the forthcoming flagship X-ray observatory \textit{Athena}. \textit{STROBE-X} too will be in low-Earth orbit and to fully leverage the capabilities of this mission to study the close environments of supermassive black holes, it will be necessary to employ a number of complementary, robust, statistical techniques to perform timing analyses spanning multiple observational gaps, of which this framework is one.

\section*{Acknowledgements}
DRW is supported by NASA through Einstein Postdoctoral Fellowship grant number PF6-170160, awarded by the \textit{Chandra} X-ray Center, operated by the Smithsonian Astrophysical Observatory for NASA under contract NAS8-03060. Computing for this project was performed on the Sherlock cluster. The author thanks Stanford University and the Stanford Research Computing Center for providing computational resources and support. This work uses observations obtained with \textit{XMM-Newton}, an ESA science mission with instruments and contributions directly funded by ESA Member States and NASA. The author thanks the anonymous referee for their feedback on the original version of this manuscript.

\bibliographystyle{mnras}
\bibliography{agn}

\label{lastpage}

\end{document}